\documentclass[11pt,a4paper]{article}
\pdfoutput=1
\usepackage{jheppub}

\usepackage{color}
\usepackage{amsmath}
\usepackage{pifont}   
\usepackage{verbatim}       
\usepackage{acronym} 
    
\usepackage{amsfonts}
\usepackage{amssymb}
\usepackage{mathrsfs} 
\usepackage{graphicx}
\usepackage{multirow} 
\usepackage{slashed} 
\usepackage{array}

\usepackage{graphicx}
\usepackage{epsfig}
\usepackage{lscape}
\usepackage{rotating}
\usepackage{epsf}
\usepackage{bm}
\usepackage{booktabs}
\usepackage{array}
\usepackage{caption}
\usepackage{subcaption}
\usepackage[utf8]{inputenc}
\usepackage{float}

\usepackage{algorithm}
\usepackage{algpseudocode}
\usepackage{afterpage}
\usepackage{multirow}
\usepackage{array,booktabs,colortbl,multirow}
\usepackage{colortbl,xcolor,colordvi,color}
\usepackage{rotating}

\usepackage{appendix}

\allowdisplaybreaks

\title{Improving the performance of weak supervision searches using transfer and meta-learning}

\author[a]{Hugues Beauchesne,}
\author[b]{Zong-En Chen}
\author[b,a]{and Cheng-Wei Chiang}

\affiliation[a]{Physics Division, National Center for Theoretical Sciences,\\ Taipei 10617, Taiwan}
\affiliation[b]{Department of Physics and Center for Theoretical Physics, National Taiwan University, \\ Taipei 10617, Taiwan}

\emailAdd{beauchesneh@phys.ncts.ntu.edu.tw, r10222045@ntu.edu.tw, chengwei@phys.ntu.edu.tw}

\abstract{Weak supervision searches have in principle the advantages of both being able to train on experimental data and being able to learn distinctive signal properties. However, the practical applicability of such searches is limited by the fact that successfully training a neural network via weak supervision can require a large amount of signal. In this work, we seek to create neural networks that can learn from less experimental signal by using transfer and meta-learning. The general idea is to first train a neural network on simulations, thereby learning concepts that can be reused or becoming a more efficient learner. The neural network would then be trained on experimental data and should require less signal because of its previous training. We find that transfer and meta-learning can substantially improve the performance of weak supervision searches.  }

\begin{document}

\maketitle

\section{Introduction}\label{Sec:Intro}

The recent advances in deep learning have created many opportunities for collider physics. A potential albeit obvious application of deep learning is to help discriminate a signal from background and thus possibly discover new particles. To be able to do so, a neural network (NN) needs to be trained. How this training is performed and which kind of training data to use are non-trivial questions and many strategies have been proposed.

Fully supervised learning, where all training data are labelled, is one such potential strategy and was considered in, e.g., Refs.~\cite{Chen:2019uar, Bernreuther:2020vhm, Chang:2020rtc, Chiang:2022lsn}. However, since the goal would be to find a signal that has not been observed yet, training must be performed based upon simulations. There are two potential risks that arise from this. First, simulations unavoidably contain artefacts. This can result in the neural network learning from these artefacts and behaving in a suboptimal and difficult-to-predict way on real data~\cite{Metodiev:2017vrx}. Second, how the neural network would react to a signal that differs from the expected one is unclear. This can potentially result in a search that is only sensitive to a very narrow range of models and thus possibly missing a discoverable signal.

Unsupervised learning, where the training data are not labelled, is another potential training strategy. Up to now, the most common approach has been to train an autoencoder on what is presumably mostly background events and use the reconstruction error as a test statistic~\cite{Farina:2018fyg}. There are again two potential drawbacks to this. First, the reconstruction error has been shown to sometimes be a very poor discriminating factor~\cite{Batson:2021agz}. Second, being trained on background only, autoencoders cannot learn any special properties the signal may possess and thus lose some of their potential discriminative power. Alternative unsupervised learning approaches exist~\cite{DAgnolo:2018cun, Nachman:2020lpy, Andreassen:2020nkr, Hallin:2021wme, Kasieczka:2022naq, Hallin:2022eoq}, but by their nature will be subject to at least the second drawback.

Weakly supervised learning, where the training data only contain imperfect labels, is a training strategy that potentially circumvents the problems of both fully supervised and unsupervised learning. In the Classification Without Label (CWoLa) method~\cite{Metodiev:2017vrx}, two samples of experimental data with presumably different fractions of signal and background are considered. Assuming the properties of the signal and background to be the same between the two samples, it is a theorem that the most powerful test statistic to distinguish the two samples is also the most powerful test statistic to distinguish the signal from the background. As such, a neural network that has been optimally trained to tell the two samples apart would also be optimally trained to tell the signal apart from the background. This neural network could then be used to suppress the background. It would also be a neural network trained specifically for the signal present in the data. As such, it would not suffer from the problem of the training and actual signal being different, unlike full supervision. This approach would combine the advantages of both unsupervised learning, i.e., being trained on data, and supervised learning, i.e., learning to exploit the special properties of the signal. An experimental search using CWoLa was performed in Ref.~\cite{ATLAS:2020iwa}.

However, there are some practical limitations associated with weak supervision that come from the fact that the amount of signal is limited~\cite{Collins:2021nxn}. In practice, there is a threshold in the amount of signal below which the neural network fails to learn properly. If not careful, this threshold can be close or above what would be sufficient for discovery even without the network. This greatly reduces the usefulness of such neural networks. This can happen for example when the input of the neural network is too large~\cite{Dillon:2023zac} and has been circumvented in Ref.~\cite{Collins:2018epr} by providing a very simple but limiting input to the network. Recent attempts at overcoming this problem include Refs.~\cite{Finke:2023ltw, Freytsis:2023cjr}.

What seems to be fundamentally the problem and which we will illustrate abundantly throughout this work is that a neural network can require a very large amount of training data to properly learn a task. In the context of a search, the small amount of signal limits the ability of the neural network to learn and decreases the effectiveness of the CWoLa method. What is ultimately the goal is then to create a neural network that requires less data to learn a task. This way, the learning threshold would be lowered and the overall performance of the CWoLa method would improve.

The strategy that we propose in this work is to use simulations to create neural networks that can learn faster from the actual data. In broad terms, the amount of signal is limited, but it is generally easy to produce simulations of it. A neural network would then first be trained on simulations to either learn certain concepts or become a more efficient learner. After this, the neural network would be trained on the data itself. It should then require less signal to learn because of what it has already learned from the simulations. In more precise terms, we will consider the use of transfer and meta-learning.

The general idea of transfer learning is for a neural network to acquire knowledge from a previous task and apply it to learning a new task~\cite{5288526, Pratt1991DirectTO}. While training on the new task, the neural network will be able to reuse previous knowledge without having to reacquire it and should therefore learn faster. In the context of a search, the neural network might learn from simulations concepts like multiplicity or thrust and then reuse these concepts once it starts learning from the actual data. Before being exposed to the data, the neural network would not know the expected distributions of multiplicity or thrust, but it would at least know what they are and that they are potentially useful observables. In practice, we will use pretraining, which will be more carefully explained in Sec.~\ref{Sec:TransferLearning}.

The general idea of meta-learning is not so much to learn a given task, but more to create a better learner. This is why it is sometimes referred to as learning-to-learn (see Ref.~\cite{9428530} for a useful review). The general approach is to first submit the neural network to a phase of meta-training. During this process, the neural network learns multiple tasks and some adjustments are performed to make the neural network learn new tasks increasingly faster. Hopefully, the neural network should learn faster once trained on the actual data. The different meta-learning techniques fall into three categories: Optimization-based, model-based and metric-based. In practice, the technique we will explore is meta-transfer learning (MTL)~\cite{sun2019meta}, which will be more carefully explained in Sec.~\ref{Sec:Meta-Learning}.\footnote{See Ref.~\cite{Dolan:2021pml} for another application of meta-learning to collider physics.}

As a benchmark, we will use so-called dark showers. These are jets that originate from new confining dark sectors (see Ref.~\cite{Albouy:2022cin} for a review). They are common in various solutions to the hierarchy problem (see, e.g., Refs.~\cite{Chacko:2005pe, Graham:2015cka}), can provide many potential dark matter candidates (see, e.g., Refs.~\cite{Beauchesne:2018myj, Bernreuther:2019pfb, Beauchesne:2019ato}), and have been the subject of several experimental searches~\cite{CMS:2018bvr, CMS:2021dzg, ATLAS:2023swa, ATLAS:2023kao}. They are also a prime target for weak supervision, as they could take many forms and our ability to simulate them accurately is still not firmly established~\cite{Bardhan:2023mia}. Most importantly, the {\tt Pythia}~\cite{Bierlich:2022pfr} Hidden Valley (HV) module~\cite{Carloni:2011kk, Carloni:2010tw} offers great flexibility in the choice of parameters, which will enable learning from a wide scope of signals.

We find the following results. Transfer learning can substantially improve the performance of CWoLa searches. The improvement is most drastic at low significance and the amount of signal necessary for discovery can sometimes be several times smaller. Meta-transfer learning can further enhance the performance of CWoLa searches, but the improvement is less than between regular CWoLa and transfer learning.

The paper is organized as follows. The event generation is presented in Sec.~\ref{Sec:EventGeneration}. Sec.~\ref{Sec:CWoLa} reviews the CWoLa method. Transfer learning is discussed in Sec.~\ref{Sec:TransferLearning} and meta-learning in Sec.~\ref{Sec:Meta-Learning}. Some concluding remarks are presented in Sec.~\ref{Sec:Conclusion}. The potential impact of systematic uncertainties is discussed in Appendix~\ref{appendix:systematic-uncertainty}.

\section{Events generation}\label{Sec:EventGeneration}

The {\tt Pythia} Hidden Valley module is used to simulate dark showers. The presence of many adjustable parameters allows for the generation of a wide range of signals, which makes this module especially convenient for transfer and meta-learning. More precisely, the signal considered is $p p \to Z' \to \bar{q}_D q_D$. The dark quarks $q_D$ are a set of new fermions charged under a new confining gauge group but neutral under the Standard Model (SM) gauge groups and assumed to be degenerate in mass. The particle $Z'$ is a massive Abelian gauge boson that interacts with both quarks and dark quarks. The resulting signature is a pair of dark jets with an invariant mass consistent with $Z'$.

Once produced, the dark quarks are showered and hadronized by {\tt Pythia}~8.307. The resulting dark hadrons are either vector mesons $\rho_D$ or pseudo-scalar mesons $\pi_D$. The ratio of their masses is set following the recommendations of Ref.~\cite{Albouy:2022cin}:
\begin{equation}
\label{Eq:relation of mass}
    \frac{m_{\pi_D}}{\Lambda_D}=5.5\sqrt{\frac{m_{q_D}}{\Lambda_D}}, 
    \quad 
    \frac{m_{\rho_D}}{\Lambda_D}=\sqrt{5.76+1.5\frac{m_{\pi_D}^2}{\Lambda_D^2}}, 
    \quad 
    m_{q_{\rm const}}=m_{q_D}+\Lambda_D,
\end{equation}
where $m_{q_D}$ and $m_{q_{\rm const}}$ are the current and constituent mass of the dark quarks respectively and $\Lambda_D$ is the dark confining scale. Note that the dark quark mass in the HV settings of {\tt Pythia} is the constituent mass. The decay of $\rho_D \to \pi_D \pi_D$ is allowed if ${m_{\pi_D}}/{\Lambda_D}<1.52$.

Two scenarios are considered for the decay of $\rho_D$. In the first scenario, $m_{\rho_D}>2m_{\pi_D}$ and the decay $\rho_D \to \pi_D \pi_D$ is allowed. Since this decay is expected to dominate, the corresponding branching ratio is set to 1. We choose seven benchmarks for this scenario, each corresponding to a different $\Lambda_D$ but a constant mass ratio $m_{\pi_D}/\Lambda_D=1$. The choices of $\Lambda_D$ are 1, 5, 10, 20, 30, 40, and 50 GeV, and the corresponding $m_{\pi_D}$, $m_{\rho_D}$, $m_{q_D}$ and $m_{q_{\rm const}}$ are set by Eq.~\eqref{Eq:relation of mass} or the ratio $m_{\pi_D}/\Lambda_D$. We simply impose the dark pions to decay to the SM $d\overline{d}$. This scenario is referred to as Indirect Decay (ID).

In the second scenario, $m_{\rho_D}<2m_{\pi_D}$ and the $\rho_D \to \pi_D \pi_D$ decay is forbidden.  We also choose seven benchmarks for this scenario, each corresponding to a different $\Lambda_D$ but a constant ratio $m_{\pi_D}/\Lambda_D=1.8$. The choices of $\Lambda_D$ are the same as in the ID scenario, and the corresponding $m_{\pi_D}$, $m_{\rho_D}$, $m_{q_D}$ and $m_{q_{\rm const}}$ are set by Eq.~\eqref{Eq:relation of mass} or the ratio $m_{\pi_D}/\Lambda_D$. We also simply impose both dark pions and dark vector mesons to decay to the SM $d\overline{d}$. This scenario is referred to as Direct Decay (DD).

\begin{figure}[t!]
    \centering
    \includegraphics[width=0.75\textwidth]{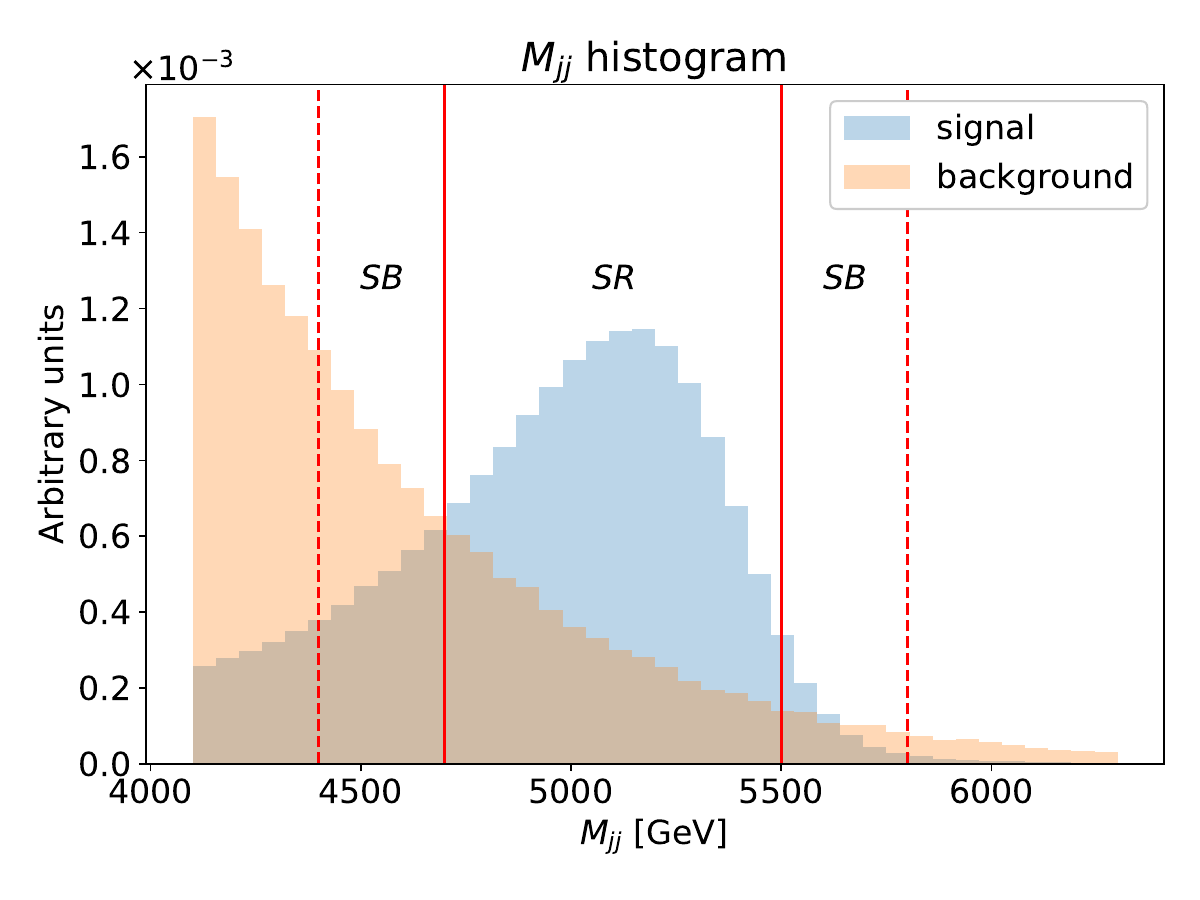}
    \caption{Dijet invariant mass distributions for the indirect decaying scenario with $\Lambda_D=10$~GeV and for the SM background. Distributions are normalized to unity. Both signal and background satisfy the selection criteria of Table~\ref{tab:madgraph-macro} except for the SR or SB conditions.}
    \label{fig:Massdistribution}
\end{figure}

The other relevant signal parameters are as follows. The mass of $Z'$ is set to 5.5 TeV, which leads to an invariant mass of the leading two jets of around 5.2 TeV. The slight difference is due to some constituents falling outside the reconstructed jets. Fig.~\ref{fig:Massdistribution} shows the distribution of the invariant mass of the two leading jets $M_{jj}$. The width of $Z'$ is taken as 10~GeV, which does not lead to any sizeable peak widening that could adversely affect the search. The values of the other HV parameters are shown in Table~\ref{tab:all parameter}.

\begin{table}[t!]
\begin{subtable}[h]{0.45\textwidth}
    \centering
    \begin{tabular}{|l|l|}
    \hline
    HV parameters in {\tt Pythia} & \\
    \hline
     HiddenValley: alphaOrder & 1\\
     HiddenValley: nFlav  &3\\
     HiddenValley: Ngauge & 3\\
     HiddenValley: pTminFSR & $1.1\Lambda_D$\\
     HiddenValley: separateFlav & on\\
     HiddenValley: aLund & 0.1\\
     HiddenValley: bmqv2 & 1.9\\
     HiddenValley: rFactqv &1.0\\
     HiddenValley: probVector & 0.75\\
     HiddenValley: fragment & on\\
     HiddenValley: FSR & on\\
     \hline
    \end{tabular}
    \caption{}
    \label{tab:all parameter}
\end{subtable}
\begin{subtable}[h]{0.6\textwidth}
    \centering
    \begin{tabular}{|l|}
    \hline
    Preliminary cuts in {\tt Madgraph}\\
    \hline
    $\sqrt{s}=13$ TeV\\
    Both $P_T$ of the leading two jets  $> 700 $~GeV \\
     Both $\eta$ of leading two jets $|\eta_j|< 2.2$ \\
     $M_{jj}>3000 $GeV \\
     \hline
     \hline
     Selection criteria after {\tt Delphes}\\
     \hline
      Number of jets $n_j\geq 2$\\
    Both $P_{T}$ of the leading two jets $> 750 $~GeV  \\
     Both $\eta$ of leading two jets $|\eta_j|< 2$ \\
      SR=$\{M_{jj}\in[4700,5500] \}$ \\
      SB=$\{M_{jj}\in[4400,4700]\cup [5500,5800] \}$ \\
     \hline
    \end{tabular}
    \caption{}
    \label{tab:madgraph-macro}
\end{subtable}
 \caption{(a) Parameters for dark showering in {\tt Pythia}. (b) Parameters in {\tt Madgraph} and the selection criteria after {\tt Delphes}.}
\end{table}

The dominant background is expected to be pair production of QCD jets. Background events are generated at parton level using {\tt Madgraph}~2.7.3~\cite{Alwall:2014hca} and hadronized using {\tt Pythia}~8.307. To speed up event generations, the preliminary cuts of Table \ref{tab:madgraph-macro} are imposed in {\tt Madgraph}. It has been verified that these cuts are sufficiently weaker than the final cuts not to have any significant impact on the distributions. The parton distribution function used for both event generations is NN23LO1~\cite{Ball:2012cx}. For the background, the default {\tt Pythia} settings are used. For both signal and background, detector simulation is handled with {\tt Delphes}~3.4.2 ~\cite{deFavereau:2013fsa}. The default CMS card is used, except for the jet radius which is set to $R=0.8$. After detector simulations, we impose the selection criteria described in Table~\ref{tab:madgraph-macro}. A Signal Region (SR) and Sidebands (SB) are defined and will come into play in the CWoLa procedure.

\begin{figure}[t!]
    \centering
    \begin{subfigure}{0.49\textwidth}
        \centering
        \includegraphics[width=\textwidth]{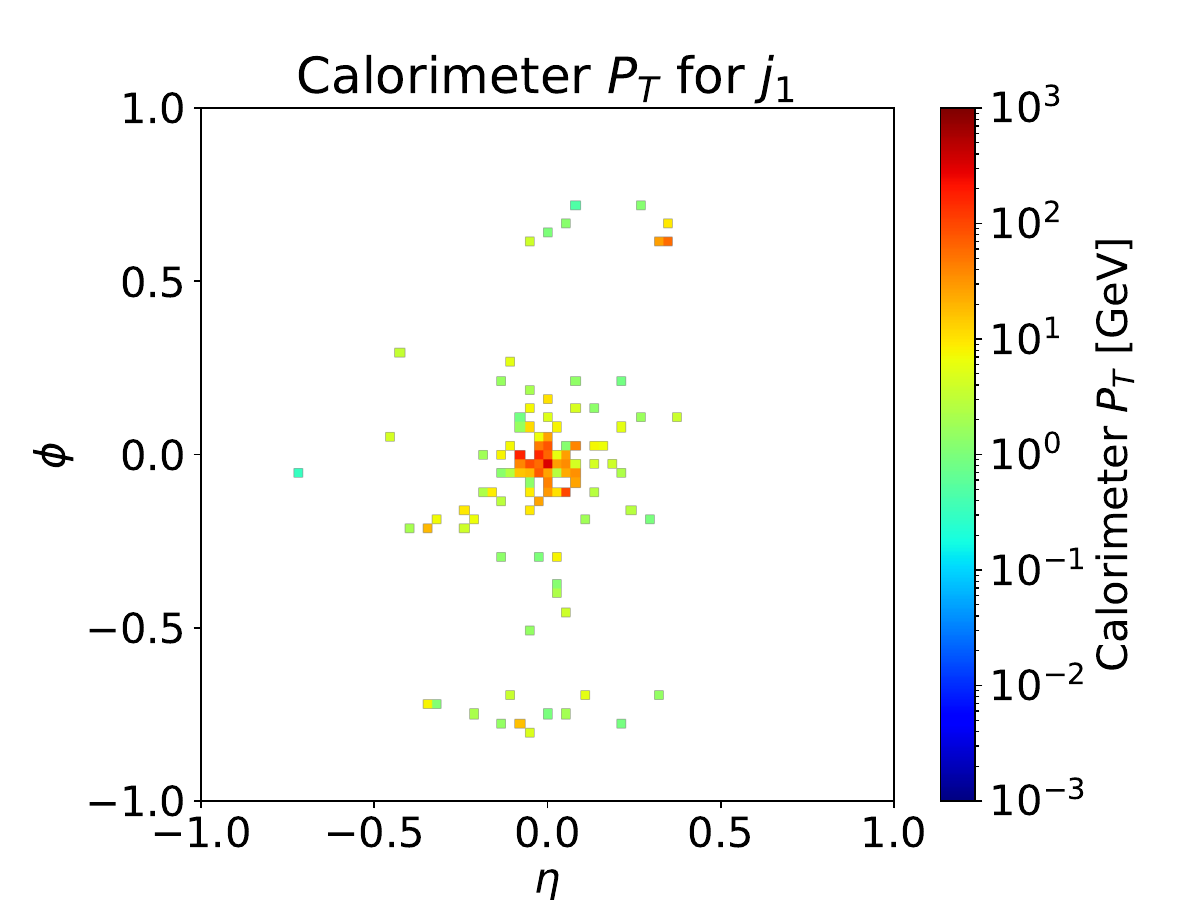}
        \caption{Before preprocessing}
        \label{subfig:before-preprocessing}
    \end{subfigure} 
    \begin{subfigure}{0.49\textwidth}
        \centering
        \includegraphics[width=\textwidth]{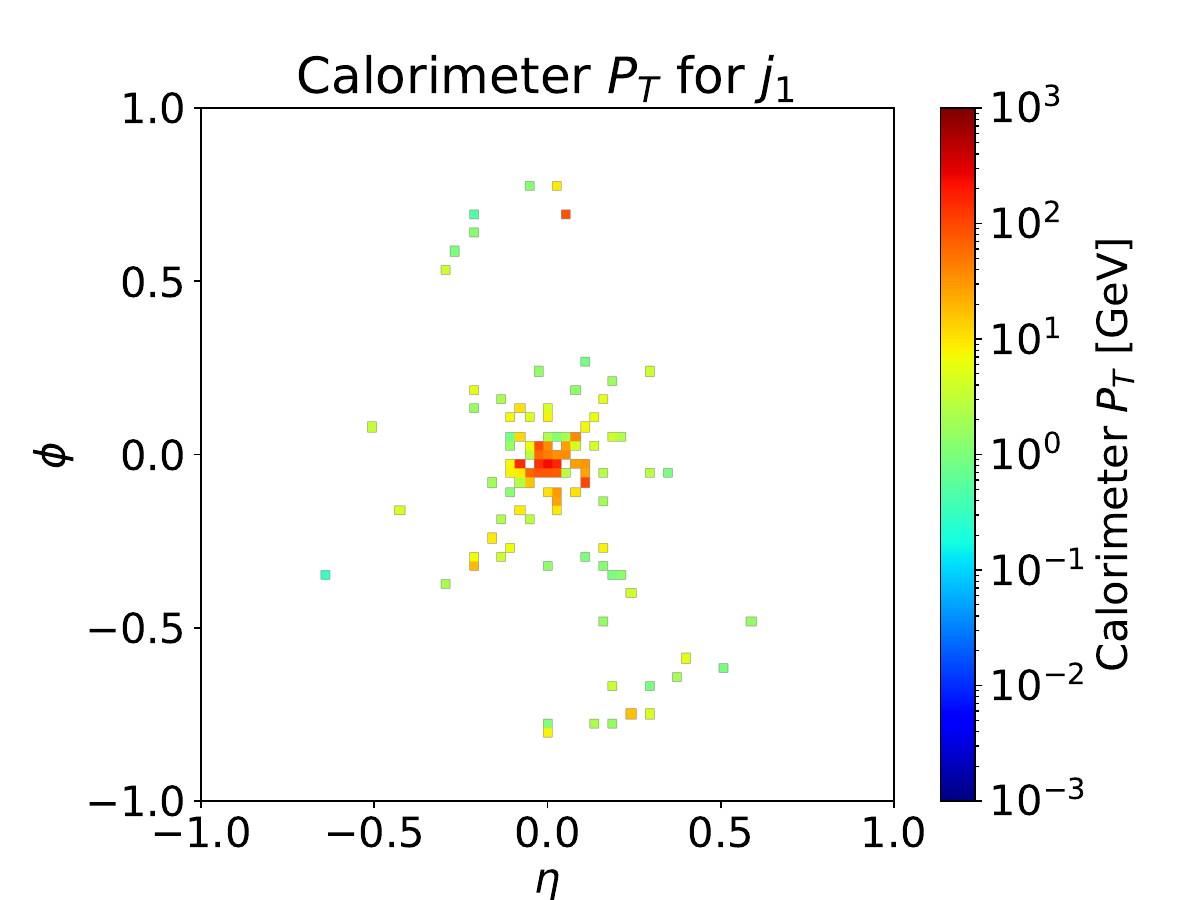}
        \caption{After preprocessing}
        \label{subfig:after-preprocessing}
    \end{subfigure}    
    
    \begin{subfigure}{0.49\textwidth}
        \centering
        \includegraphics[width=\textwidth]{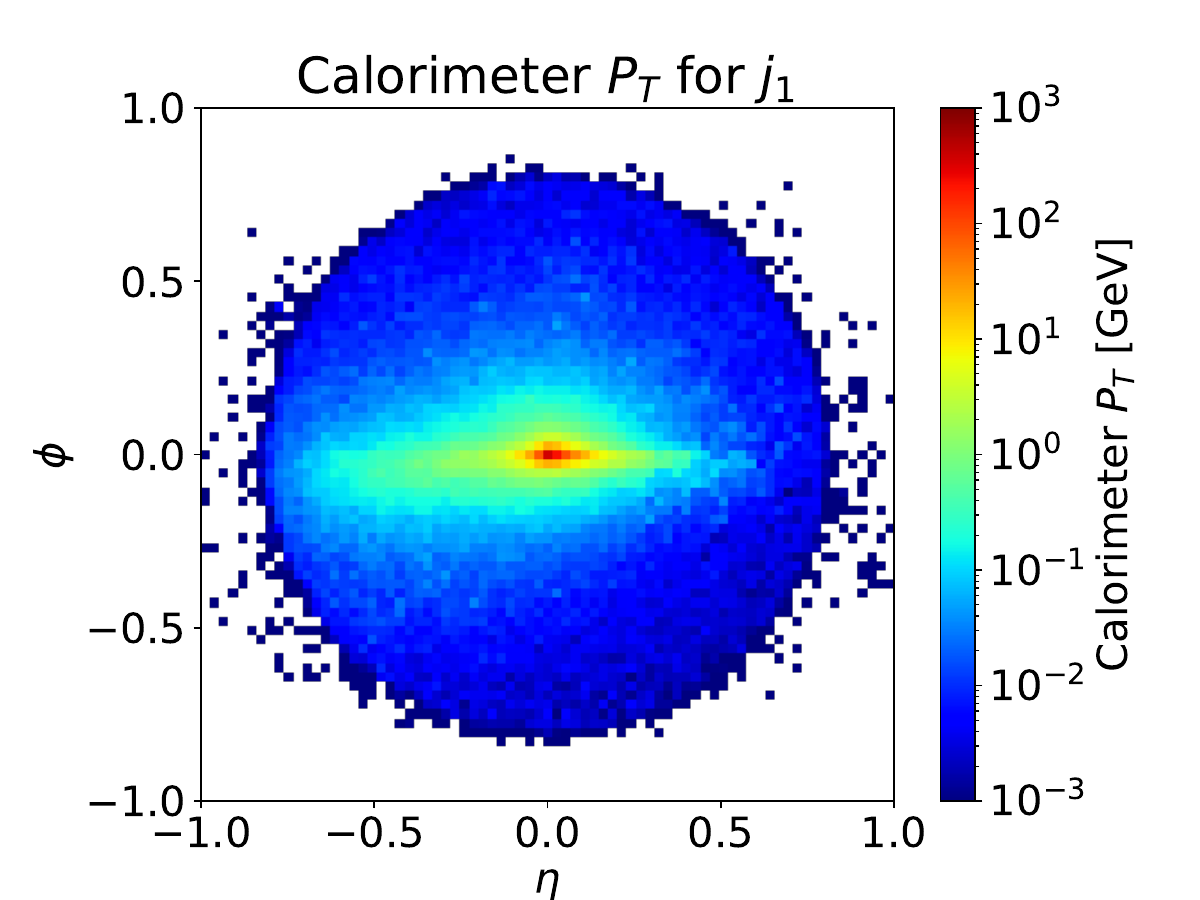}
        \caption{Average histogram of background}
        \label{subfig:bg-preprocessing}
    \end{subfigure}    
    \begin{subfigure}{0.49\textwidth}
        \centering
     \includegraphics[width=\textwidth]{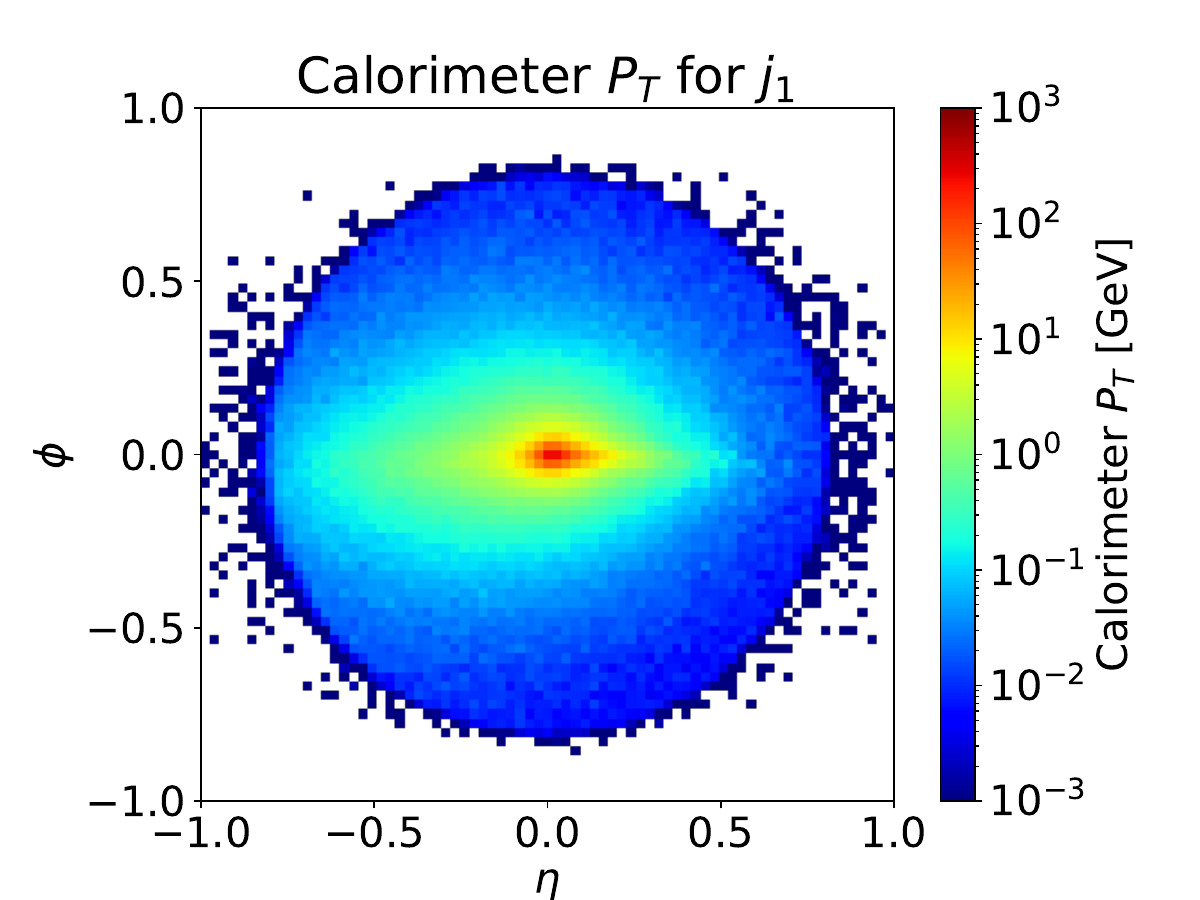}
        \caption{Average histogram of signal}
        \label{subfig:sg-preprocessing}
    \end{subfigure}    
    \caption{(a) A 2D $P_T$ histogram for one signal event in the SR before rotation and flipping. (b) A 2D $P_T$ histogram of the same event after complete preprocessing. (c) The average histogram for 10k background events in the SR after preprocessing. (d) The average histogram for 10k signal events in the SR after preprocessing. These plots are for the leading jet with $75 \times 75$ resolution and the ID scenario with $\Lambda_D$=~10~GeV.}
    \label{fig:preprocessing_plot}
\end{figure}

Finally, the two leading jets in $P_T$ are converted into jet images according to the following procedure~\cite{Kasieczka:2019dbj, deOliveira:2015xxd, Kasieczka2017nv}. First, the jet constituents are translated so that the center of the image is along the jet axis. Second, the image is rotated such that the principal axis of the $P_T$-weighted constituents is along the horizontal direction. Third, the image is flipped such that the highest $P_T$ constituent is in the upper right plane. After the above preprocessing, the image is pixelated using resolutions of either $25 \times 25$, $50 \times 50$ or $75 \times 75$. The ranges of $\eta$ and $\phi$ are both from $-1$ to $1$. Fig.~\ref{fig:preprocessing_plot} shows the jets before and after preprocessing, as well as the average histogram plots. Jet images are chosen as the input of the neural networks as learning from them can be challenging. This will display more clearly the improvements provided by transfer and meta-learning. The ability to adjust the resolution will also prove useful to illustrate certain features.

\section{CWoLa}\label{Sec:CWoLa}

As explained in the Introduction, the CWoLa method requires the existence of two mixed samples of signal and background in different proportions. A neural network is then trained to distinguish the two samples, which should hopefully result in the network learning the difference between the signal and background. In our case, a neural network is trained to distinguish the signal and sideband regions of Fig.~\ref{fig:Massdistribution}. In this section, we explain the details of our implementation of this procedure, which is partially inspired by Ref.~\cite{Collins:2018epr}.

The background in the SR consists of 25k events passing the SR selection cuts of Table~\ref{tab:madgraph-macro}.  A fifth of these are used for validation, leaving an integrated luminosity roughly corresponding to the expected number of events from Run~2 of the LHC. Considering the conceptual nature of this work, we did not implement $k$-fold cross-validation, but nothing would prevent its implementation in an actual search. The number of background events in the SBs is determined by using the same integrated luminosity as the SR. The amount of signal in the SR is varied throughout the analysis and the amount of signal in the SBs is set again by using the same integrated luminosity as the SR.  The callbacks function is used to save the best model during training by monitoring the validation loss. To test the performance of the CWoLa method, we use 20k additional signal passing the SR requirements and 20k similar background.

We use as training data the jet images of the two leading jets. The distributions of each of them are independently batched normalized. Each jet image is then passed through a common Convolutional Neural Network (CNN) subnetwork and each returns a single number. The output of the full neural network is then the product of these two numbers. The subarchitecture and training procedure are described in Table~\ref{tab:CNN_set}.
\begin{table}[t!]
    \centering
    \begin{tabular}{|c|l|}
    \hline
             &$\left(
             \begin{array}{l}
               \text{convolutional 2D layer: 64 filters with $5\times 5$ kernel size}  \\
              \text{maxpooling layer: $2 \times 2$ pool size   } 
             \end{array}
             \right)\times2$\\
               &convolutional 2D layer: 128 filters with $3\times 3$ kernel size   \\
    Layers of CNN &maxpooling layer: $2 \times 2$ pool size          \\
    subnetwork    &convolutional 2D layer: 128 filters with $3\times 3$ kernel size   \\
               &flatten layer                  \\
               &$(\text{dense layer: 128 units})\times3$ \\
               &dense layer (output): 1 unit       \\
    \hline
                  &convolutional layer padding: same\\
    Layer setting & hidden layer activation function: ReLU\\
                  & output layer activation function: Sigmoid\\
    \hline
                  & loss function: binary cross-entropy \\
                  & optimizer: Adam\\
                  & metric: accuracy\\
   Other          & batch size: 500\\
                  & learning rate: 1e-3 (base learning, pretraining)  \\
                  & learning rate: 1e-4 (CWoLa, fine-tuning, meta-learner updating) \\
                  & patience number: 20 (pretraining, meta-learning)\\
                  & patience number: 30 (CWoLa, fine-tuning)\\
    \hline
    \end{tabular}
    \caption{The CNN model subarchitecture and the hyperparameters}
    \label{tab:CNN_set}
\end{table}
All NNs are implemented using \texttt{Keras}~\cite{chollet2015keras} with \texttt{TensorFlow}~\cite{Abadi:2016kic} backend. We did investigate the possibility of using two distinct networks, but found this alternative to give typically inferior results. This seems to be caused by the lack of signal. The convolutional part of the neural network is referred to as the feature extractor and its weights and biases are collectively labelled as $\Theta$. The weights and biases of the dense layers are collectively labelled as $\theta$.

In order to evaluate the performance of the NN, we use the significance formula~\cite{ATLAS:2020yaz}
\begin{align}\label{eq:significance}
\sigma=\sqrt{2\left((N_s+N_b)\log\left(\frac{N_s}{N_b}+1\right)-N_s\right)},
\end{align}
where $N_s$ and $N_b$ are respectively the numbers of signal and background after the NN classification. We choose certain background efficiencies $\epsilon_b$ and calculate the corresponding signal efficiencies $\epsilon_s$ from the receiver operating characteristic (ROC) curve with testing data after training. It has also been verified that no significant excesses are produced via sculpting~\cite{Collins:2018epr}. The training is performed 10 times for each significance value, including sampling new events in each pseudo-experiment, and averaged. The standard deviations are computed and correspond to fluctuations from both the training and the sampling.

\begin{figure}[!ht]
    \centering
    \begin{subfigure}{0.49\textwidth}
        \centering
        \includegraphics[width=\textwidth]{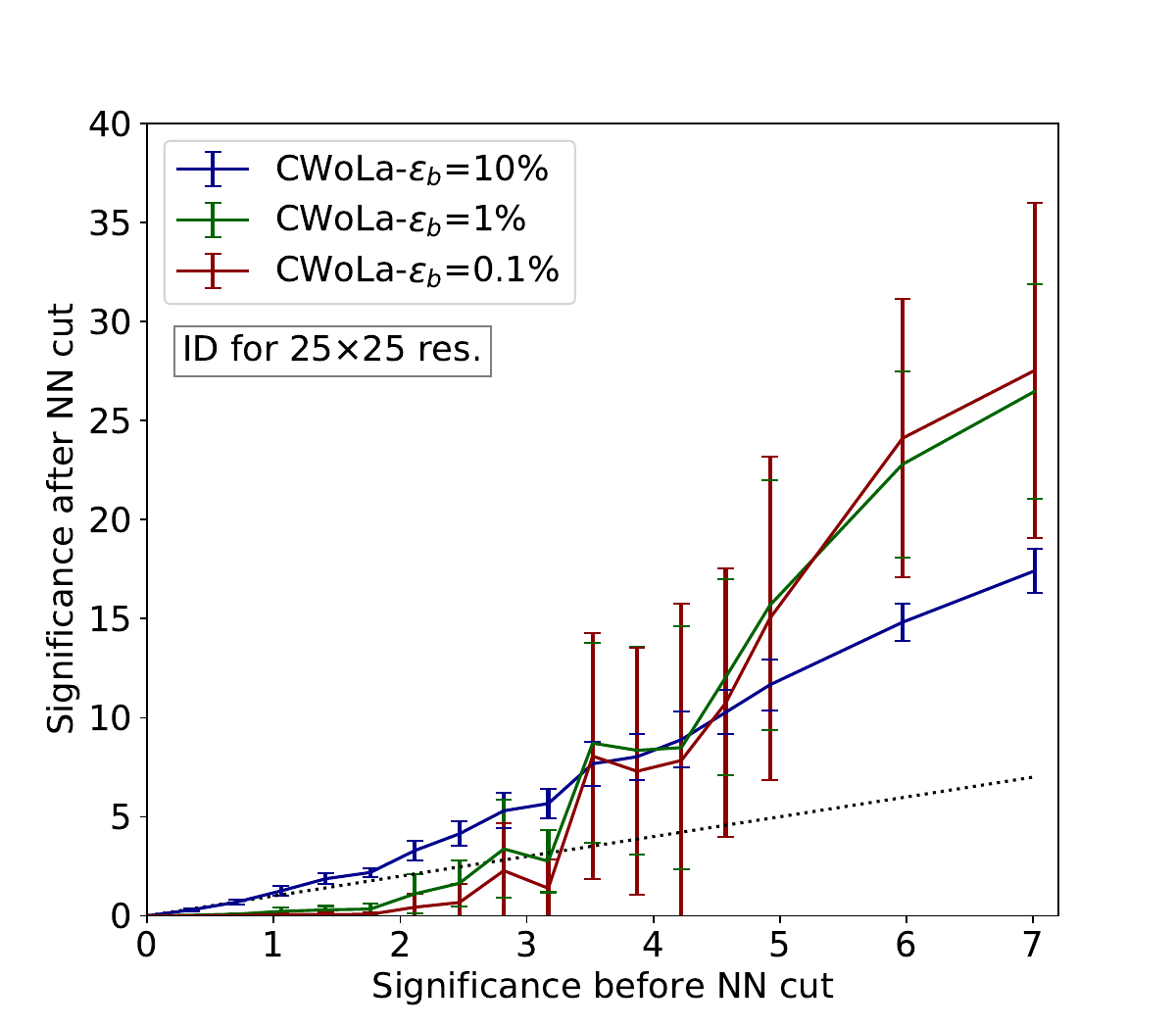}
        \caption{}
        \label{subfig:A1_CWoLa}
    \end{subfigure}    
    \begin{subfigure}{0.49\textwidth}
        \centering
        \includegraphics[width=\textwidth]{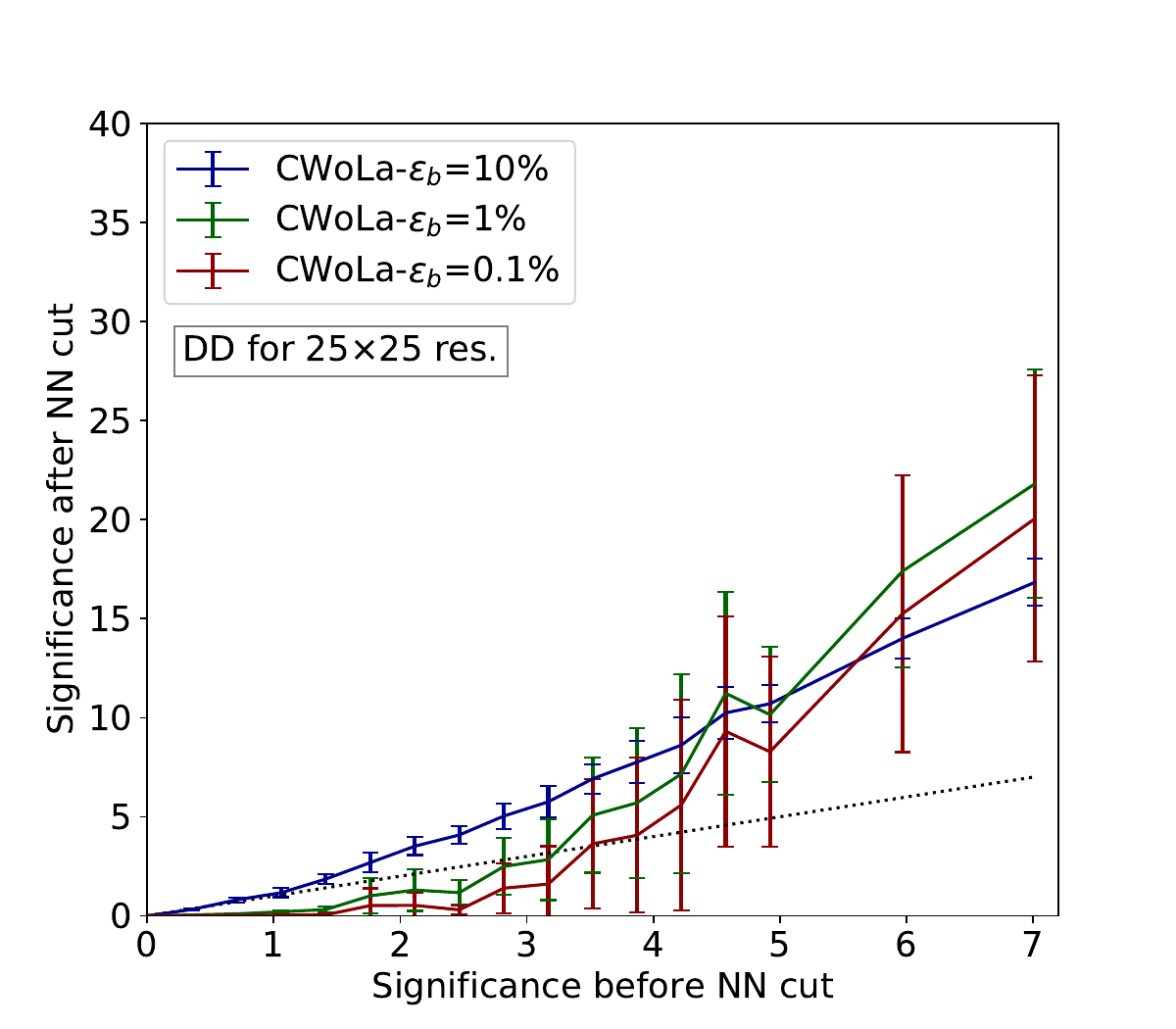}
        \caption{}
    \end{subfigure}    

    \begin{subfigure}{0.49\textwidth}
        \centering
        \includegraphics[width=\textwidth]{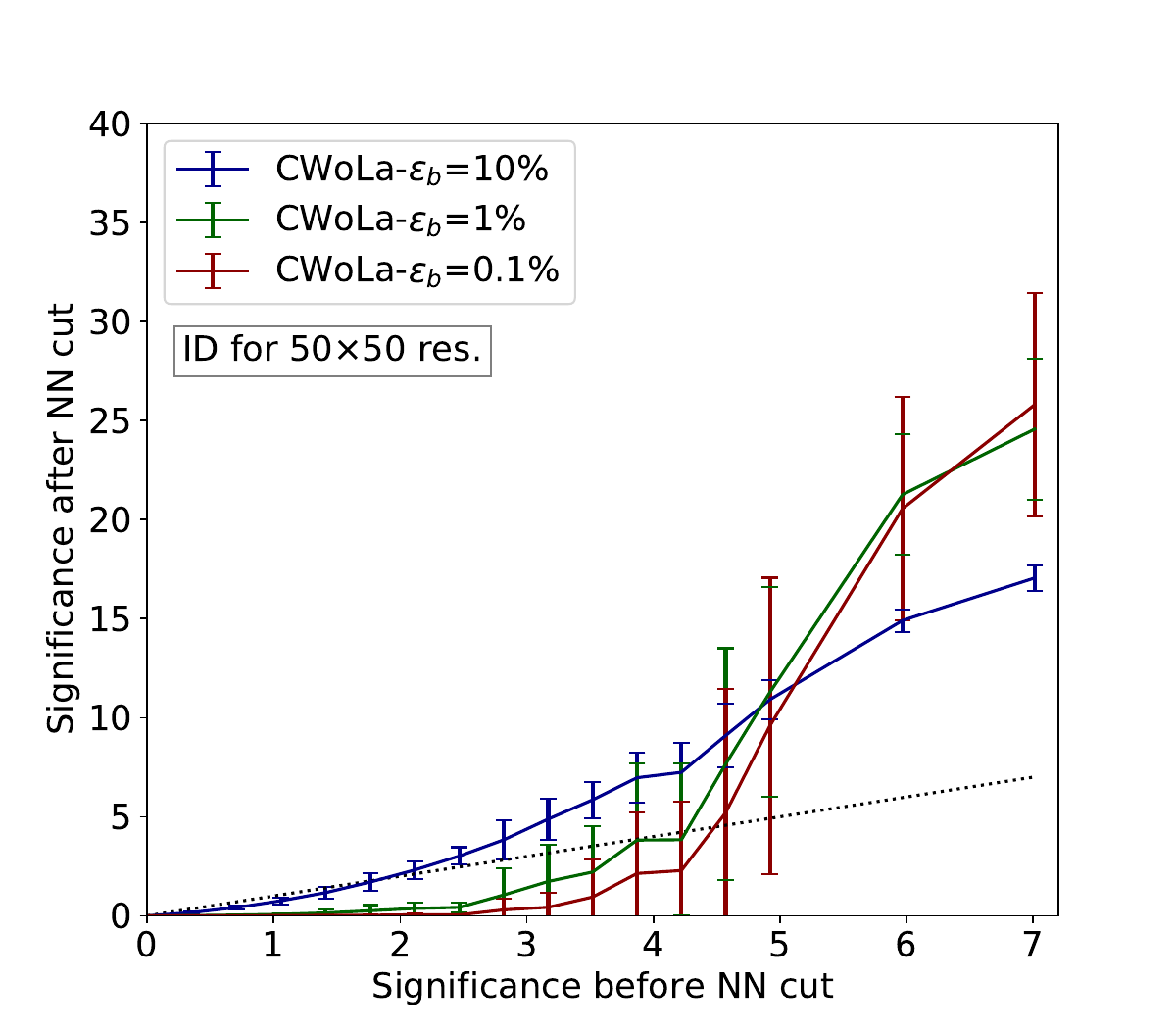}
        \caption{}
    \end{subfigure}    
    \begin{subfigure}{0.49\textwidth}
        \centering
        \includegraphics[width=\textwidth]{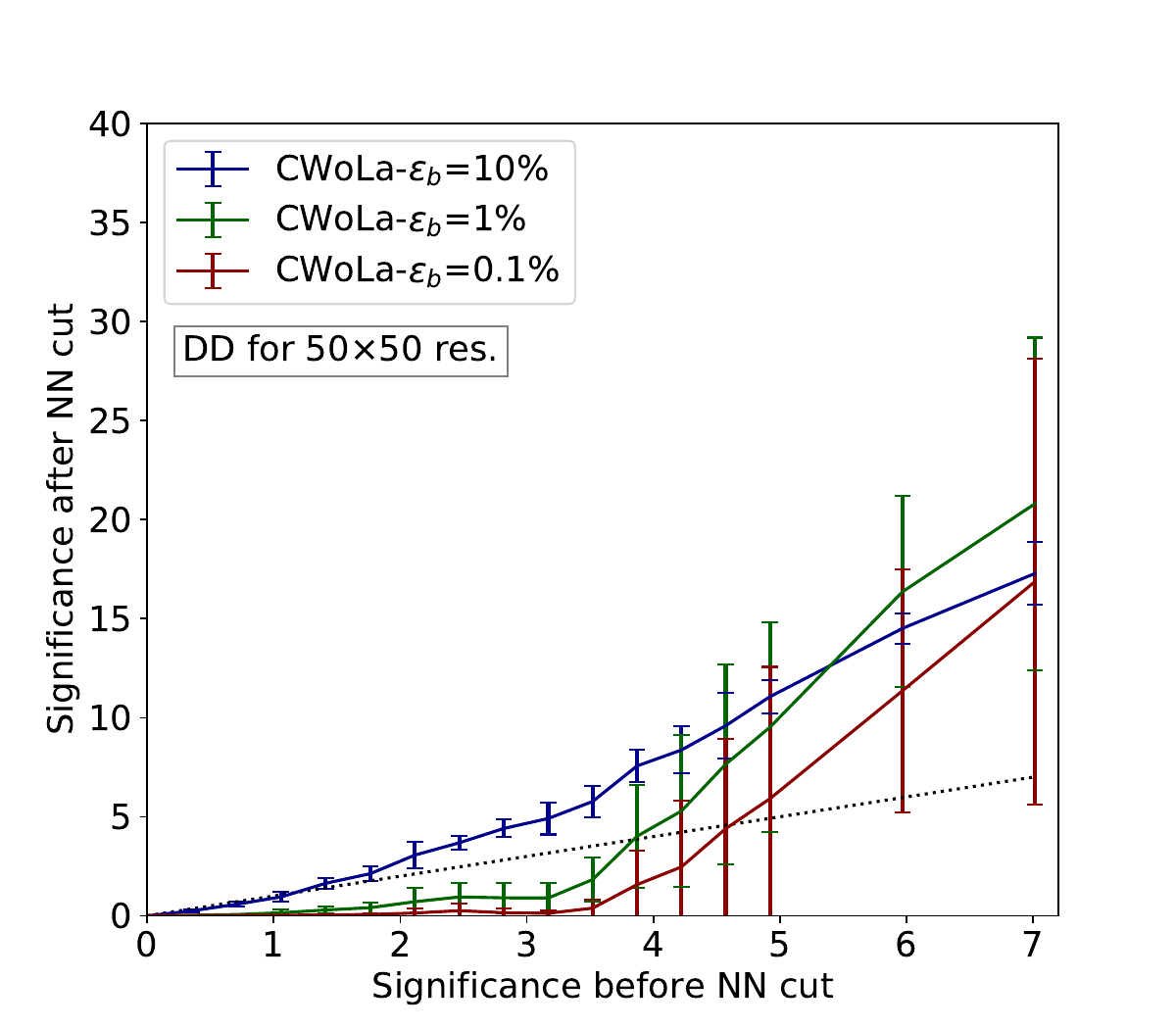}
        \caption{}
    \end{subfigure}    
    
    \begin{subfigure}{0.49\textwidth}
        \centering
        \includegraphics[width=\textwidth]{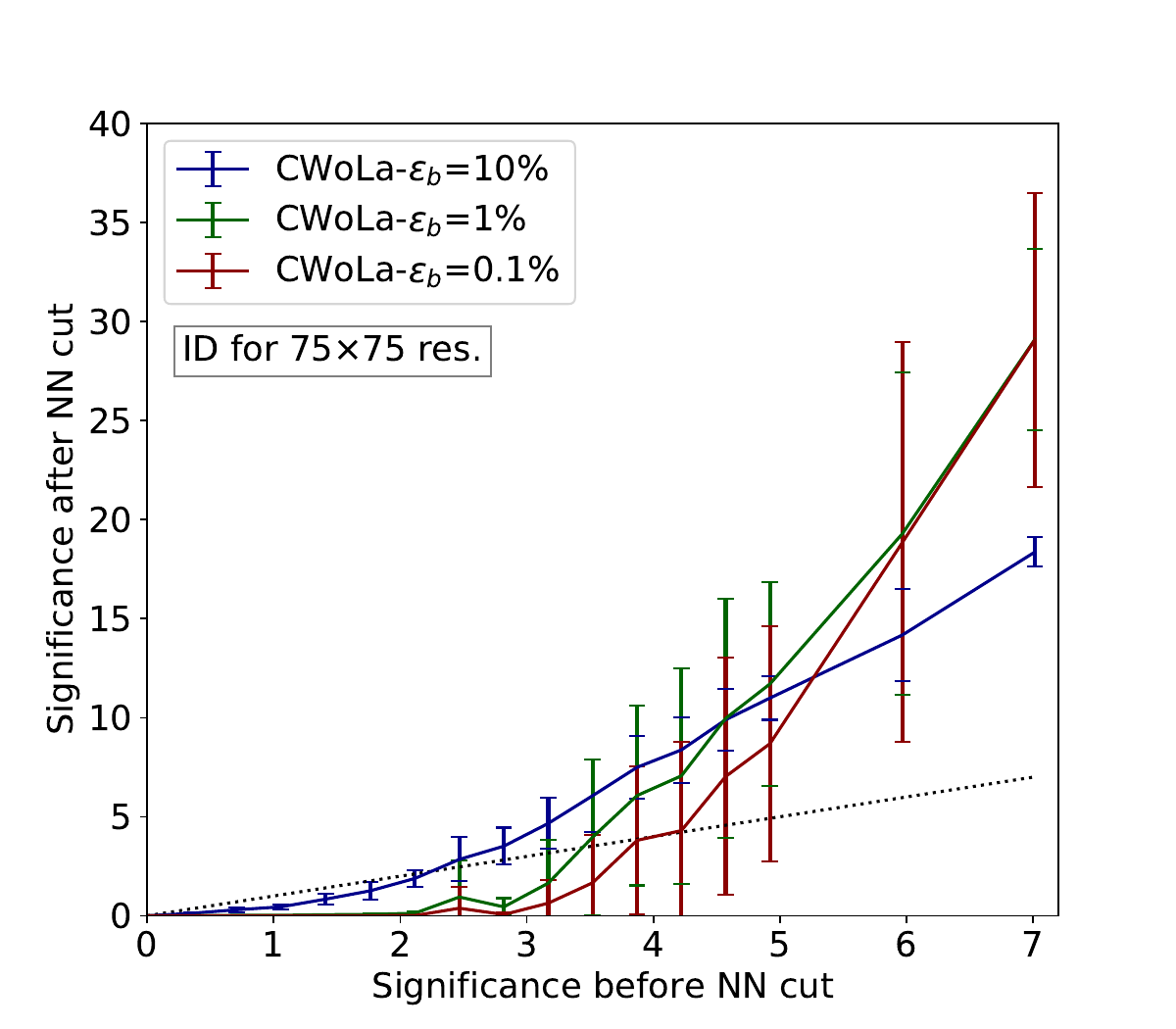}
        \caption{}
    \end{subfigure}    
    \begin{subfigure}{0.49\textwidth}
        \centering
        \includegraphics[width=\textwidth]{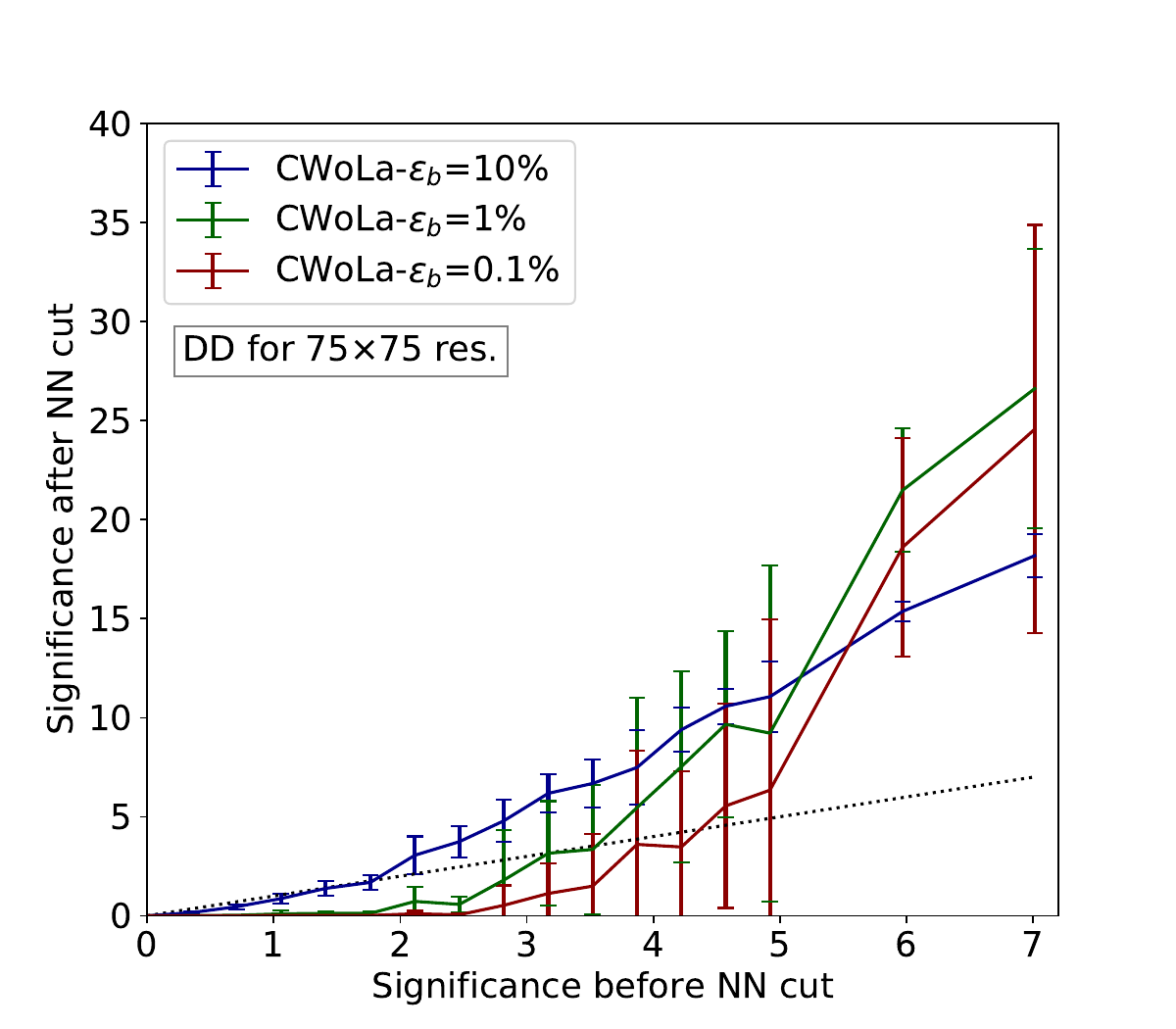}
        \caption{}
    \end{subfigure}    
    \caption{The results of CNN CWoLa for the ID (left column) and DD (right column) scenarios with $\Lambda_D$= 10~GeV for $25\times25$, $50\times50$ and $75\times75$ resolutions.  The dotted line in each plot has a slope of 1. } 
    \label{fig:result_CNN_CWoLa}
\end{figure}

Fig.~\ref{fig:result_CNN_CWoLa} shows two benchmarks with three different resolutions each. Several comments are in order.  First, the different curves display a threshold below which the neural network fails to learn from the data. This is the threshold alluded to in the Introduction and corresponds to the upward turn of the curves around 2 to 4$\sigma$. Below this threshold, the NN cuts background and signal indiscriminately and the significance is even worse than without employing the NN. Second, increasing the resolution tends to move the position of the threshold to higher significance. This is due to the fact that classifying a higher-resolution image is a more difficult task and more parameters must be learned inside the~NN.

\afterpage{\clearpage}

\section{Transfer learning}\label{Sec:TransferLearning}

As illustrated in the last section, the existence of a learning threshold renders the use of CWoLa problematic for small amounts of signal. A potential solution to this problem is transfer learning, which we introduce in this section.

The general idea of transfer learning is to have an NN first learn from a related problem with a large amount of data and then transfer some of this knowledge to the problem of interest. In practice, the technique that we use for transfer learning is pretraining, in which information is transferred via pretraining NN parameters on the larger dataset. From a terminology point of view, training an NN on a larger dataset is known as pretraining, while fine-tuning refers to the subsequent training on a smaller dataset. Furthermore, these larger and smaller datasets are respectively referred to as the source and target data.

We implement the pretraining strategy as follows. First, the NN is pretrained to distinguish a sample of pure background from a pure combination of different signals. This combination includes all the models mentioned in Sec.~\ref{Sec:EventGeneration}, except the benchmark on which the model will be tested on. In a real experiment, this would represent training on simulations.  A total of 250k signal and 250k background from the SR are used as the source data. A fifth of the sample is used for validation, which is also performed on pure samples. Second, the neural network is trained to distinguish the mixed samples, i.e., the SR and SB regions with the benchmark signal mixed within the background. In a real experiment, this would represent fine-tuning on the actual data as target data. The NN model parameters of the feature extractor $\Theta$ are initialized at their values learned during pretraining and the NN parameters of the dense layers $\theta$ are reinitialized randomly. During the fine-tuning step, $\Theta$ are frozen and only $\theta$ are trained.

\begin{figure}[t!]
    \centering
    \begin{subfigure}{0.49\textwidth}
        \centering
        \includegraphics[width=\textwidth]{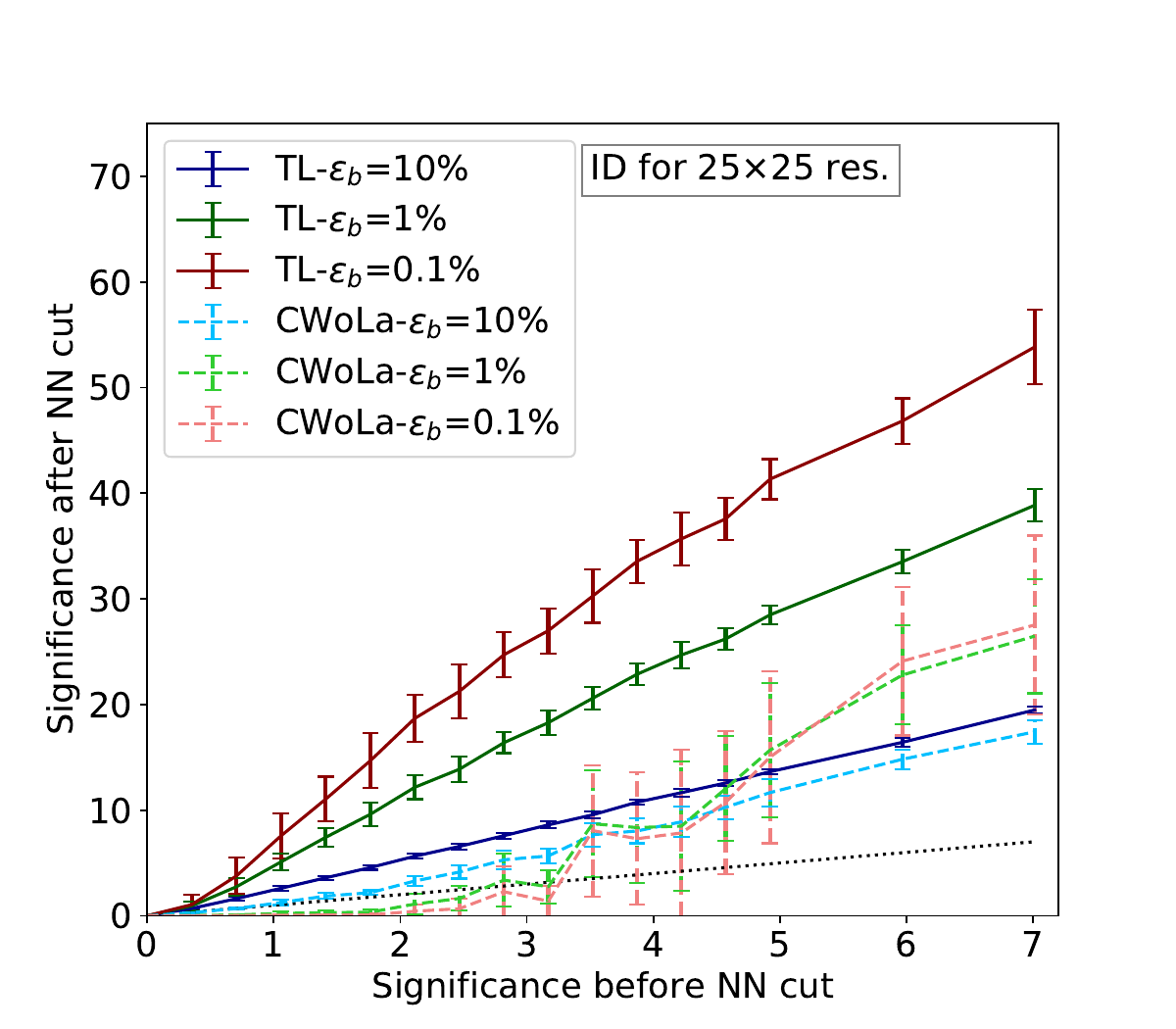}
        \caption{}
        \label{fig:TL_re25_A1}
    \end{subfigure}    
    \begin{subfigure}{0.49\textwidth}
        \centering
        \includegraphics[width=\textwidth]{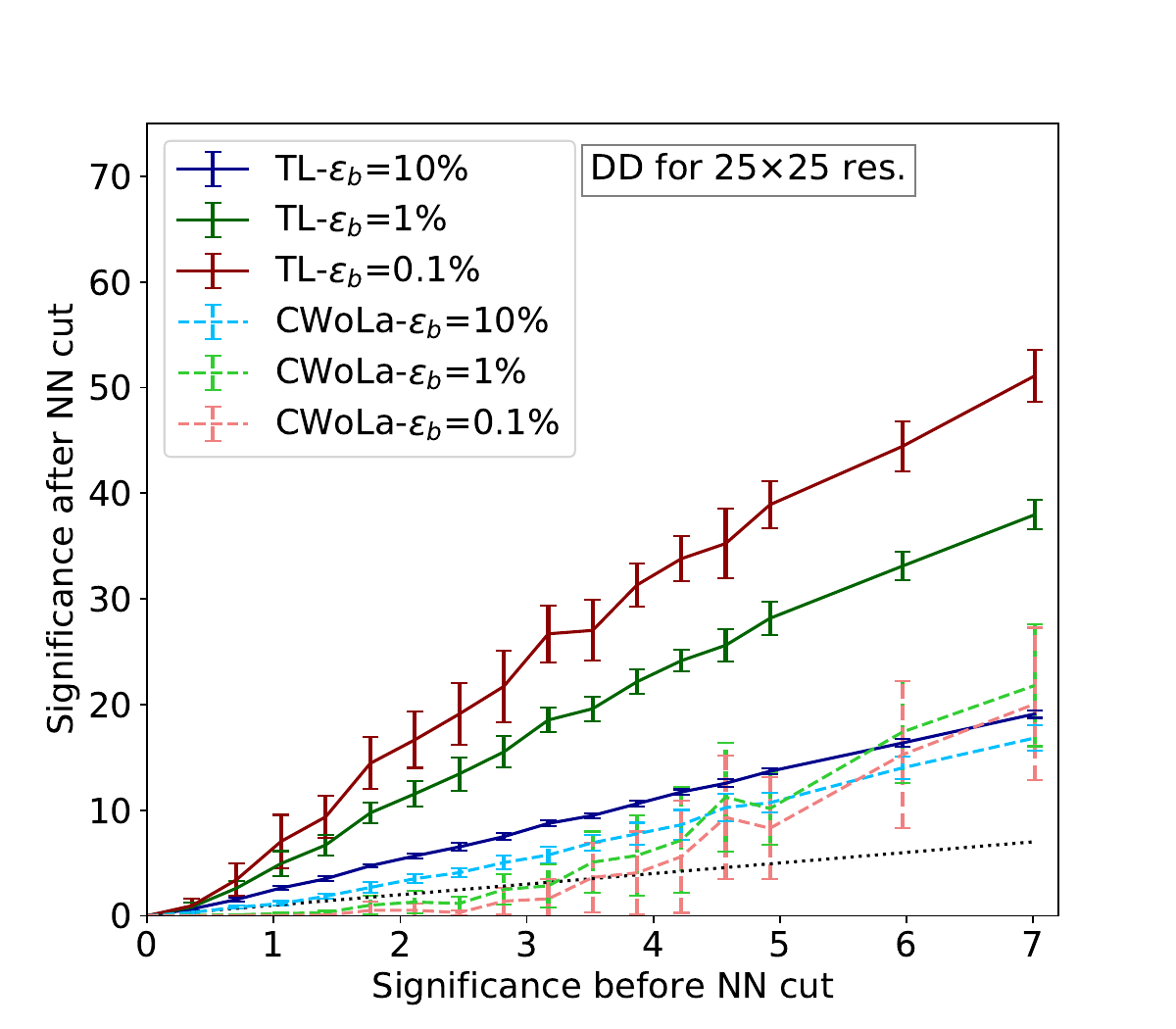}
        \caption{}
    \end{subfigure}    

    \begin{subfigure}{0.49\textwidth}
        \centering  \includegraphics[width=\textwidth]{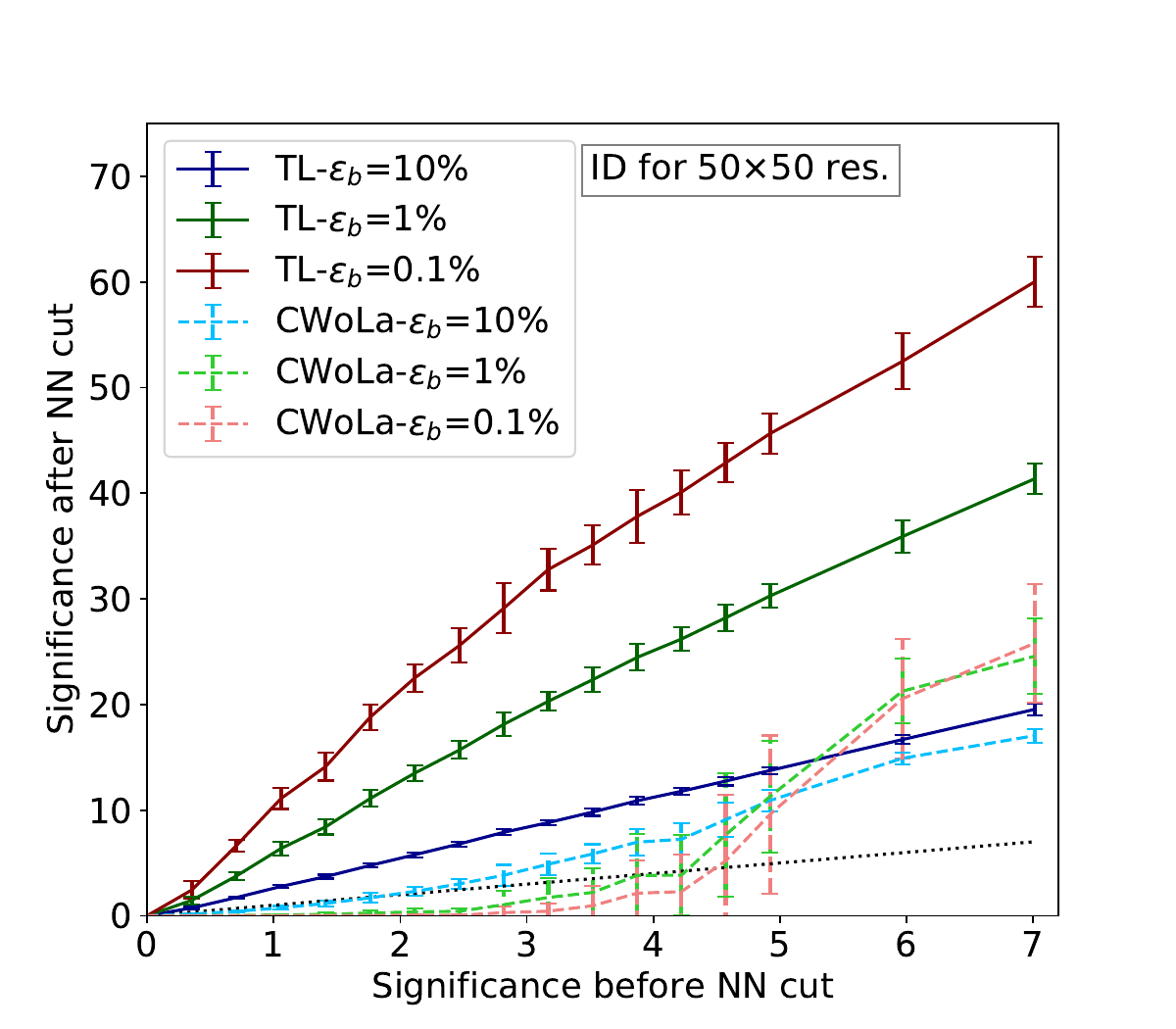}
        \caption{}
    \end{subfigure}    
    \begin{subfigure}{0.49\textwidth}
        \centering
        \includegraphics[width=\textwidth]{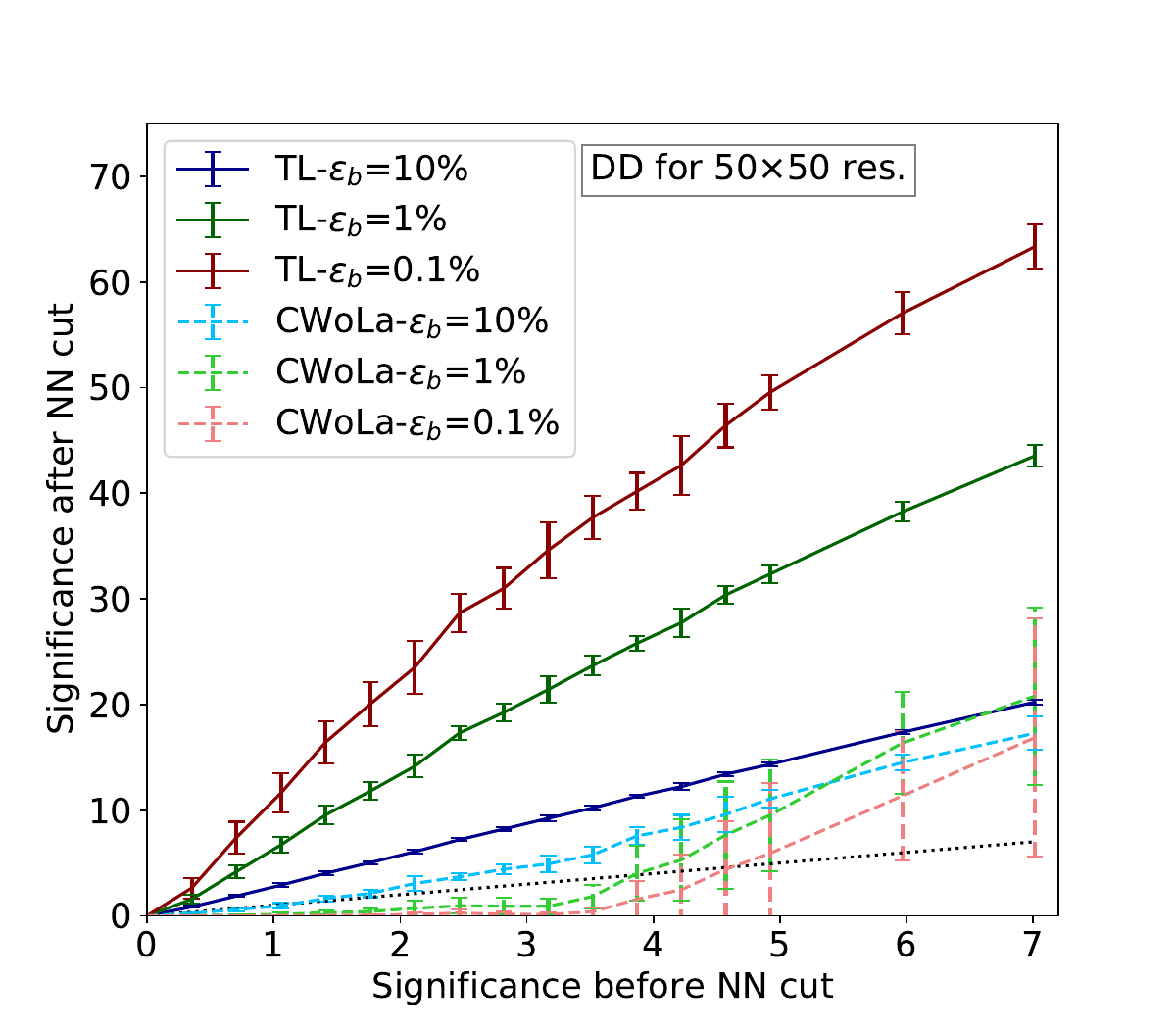}
        \caption{}
    \end{subfigure}  

    \begin{subfigure}{0.49\textwidth}
        \centering  \includegraphics[width=\textwidth]{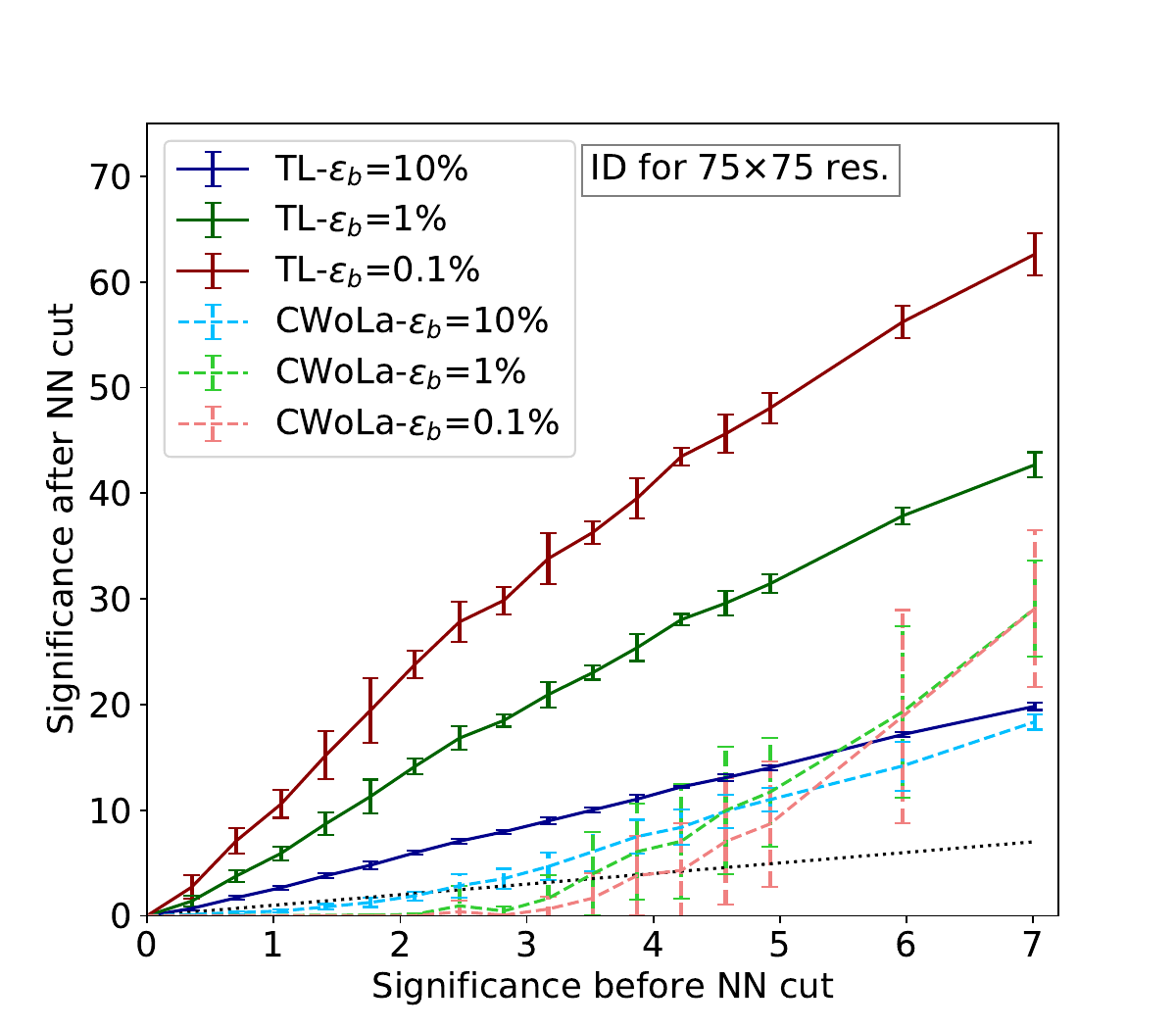}
        \caption{}
    \end{subfigure}    
    \begin{subfigure}{0.49\textwidth}
        \centering
        \includegraphics[width=\textwidth]{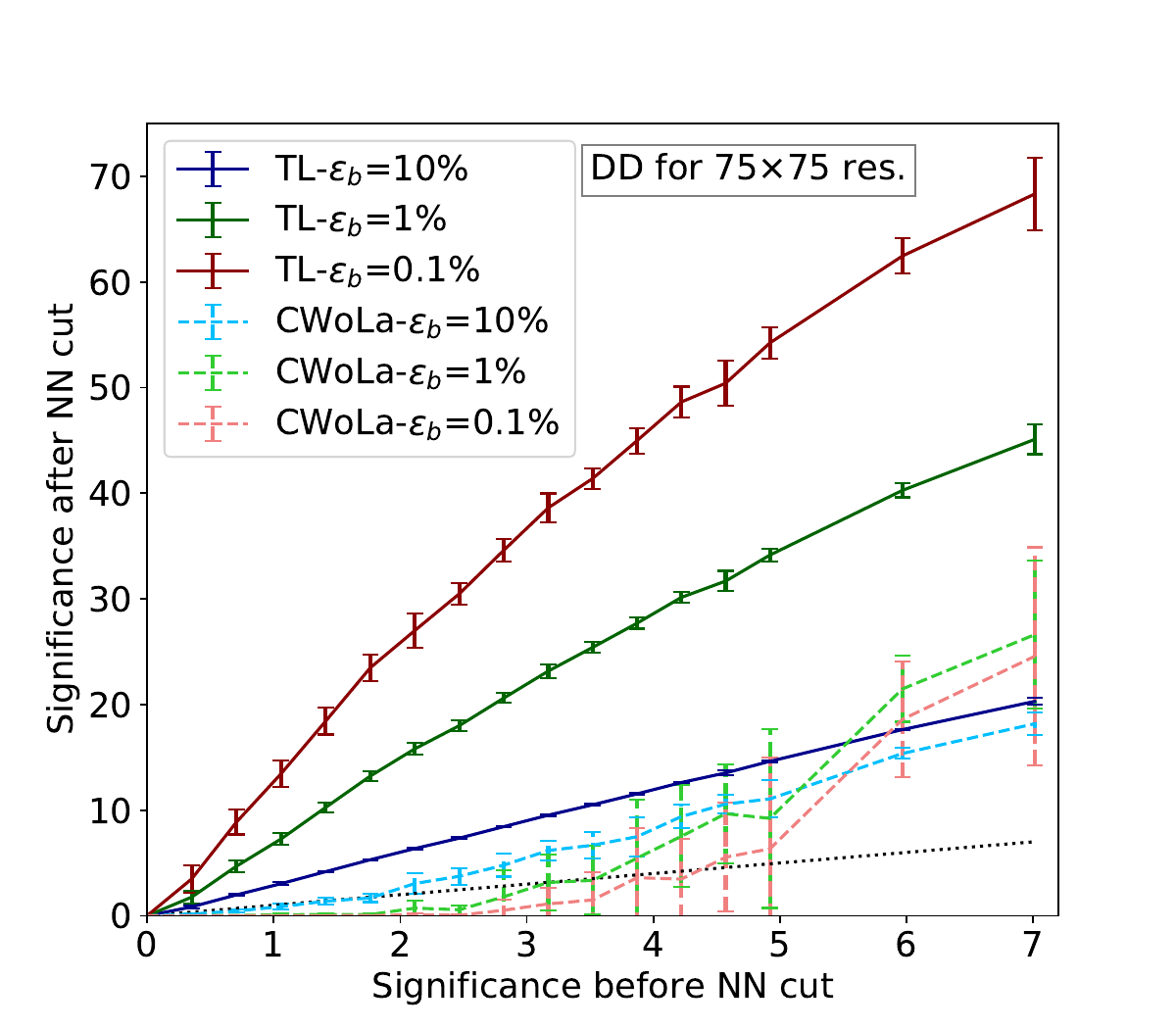}
        \caption{}
    \end{subfigure}    

    \caption{The results of transfer learning (solid curves) and of CWoLa (dashed curves, same as those in Fig.~\ref{fig:result_CNN_CWoLa}) for the ID (left column) and DD (right column) scenarios with $\Lambda_D$= 10~GeV for $25\times25$, $50\times50$ and $75\times75$ resolutions.  The dotted line in each plot has a slope of 1.}
    \label{fig:TL_CNN}
\end{figure}

Fig.~\ref{fig:TL_CNN} shows the comparison between pure CWoLa and transfer learning. First, transfer learning not only improves the general NN performance but also considerably reduces the learning threshold for all three different resolutions.  In practice, the amount of signal necessary to claim a $5\sigma$ discovery can be reduced by a factor of a few, which is due to the fact that the NN can better identify and reduce the background. Second, the relative fluctuations in the significance are reduced. This is due to a smaller amount of trainable parameters and more successful learning. Overall, the use of transfer learning displays a massive improvement over the standard CWoLa.

\afterpage{\clearpage}

\section{Meta-learning }\label{Sec:Meta-Learning}

Meta-learning is an alternative approach for creating neural networks that can learn from less data. The general idea is not so much to reuse concepts from related tasks, but more to teach the neural network how to learn tasks more efficiently. More specifically, we will study the use of meta-transfer learning (MTL)~\cite{sun2019meta}. Although many other techniques exist, we choose MTL because it is closely related to transfer learning, which has already been shown in the previous section to be very successful. We will present in this section our implementation of MTL, which we simplify and modify somewhat, and refer to Ref.~\cite{sun2019meta} for more details.

MTL uses so-called scaling and shifting parameters. Consider a rectangular image $A$ of arbitrary dimensions and $M$ channels. Assume a set of $N$ convolutional filters have previously been created. Label the filters and their indices as
\begin{equation}\label{eq:FilterDef}
  F^{cf}_{ij},
\end{equation}
where the index $f$ refers to the label of the filter and runs from 1 to $N$, the $i$ and $j$ indices correspond to the positional arguments of the filter ($\eta$ and $\phi$ in our case), and the index $c$ corresponds to the channel and runs from 1 to $M$. Scaling is then applied as
\begin{equation}\label{eq:Scaling}
  \bar{F}^{cf}_{ij} = S^{cf} F^{cf}_{ij},
\end{equation}
where $S^{cf}$ are the scaling parameters and $\bar{F}^f$ are the scaled filters. The scaled filter $\bar{F}^f$ is then applied to image $A$ at point $(i, j)$ as
\begin{equation}\label{eq:Shifting}
  B_{i'j'}^f = g\left((\bar{F}^f \star A)_{ij} + b^f + \bar{S}^f\right),
\end{equation}
where $B$ is the resulting image, $g$ is the activation function, $\star$ is the cross-correlation operation, $b^f$ are the previously determined bias and $\bar{S}^f$ are the shifting parameters. The indices $i'$ and $j'$ are related to the positions $i$ and $j$, though the exact relation depends on other parameters (stride, padding, etc). The scaling and shifting parameters are what will be optimized to make the neural network learn faster and are meant to emphasize more on important features. They are how the neural network `learns-to-learn'.

The architecture of our neural network is mostly the same as in Table~\ref{tab:CNN_set}. The only difference is the inclusion of scaling and shifting parameters in all the convolutional layers. Once again, the NN parameters of the feature extractor are referred to as $\Theta$ and those of the dense layers as $\theta$. Training proceeds in the following three phases. 

First, pretraining is performed as in Sec.~\ref{Sec:TransferLearning}, i.e., the neural network is trained to distinguish a sample of background from a mixture of different signals. The scaling parameters and the shifting parameters are kept at 1 and 0, respectively, throughout this phase. Once the pretraining is complete, the NN model parameters $\Theta$ will be fixed forever. The $\theta$ parameters are however not reinitialized randomly, which differs from the approach of Ref.~\cite{sun2019meta} but gives better results in our case.

Second, a new phase of so-called meta-training is performed. Assume a series of tasks $\mathcal{T}$ forming a task-space $p(\mathcal{T})$. For us, the tasks correspond to the different models of Sec.~\ref{Sec:EventGeneration} except the benchmark under study. The training goes schematically as follows: 

\vspace{0.2cm}
\noindent\makebox[\textwidth][c]{
\begin{minipage}{0.5\textwidth}
  \begin{algorithmic}
    \For{\texttt{episode}}
      \For{\texttt{$\mathcal{T}$ in $p(\mathcal{T})$}}
        \State \texttt{base learning}
        \State \texttt{meta-learner update}
        \State \texttt{evaluation of $\mathcal{L}_{\mathcal{T}}$}
      \EndFor
      \State \texttt{average $\mathcal{L}_\mathcal{T}$ over $p(\mathcal{T})$}
      \State \texttt{test for early stopping}
    \EndFor
  \end{algorithmic}
\end{minipage}}
\vspace{0.2cm}

\noindent In more details, an episode is the meta-learning equivalent of an epoch. In other words, each possible task in the task-space is considered during an episode and only once. The first step of every episode is an inner-loop. For each task in the task-space, the following steps are performed:
\begin{itemize}
  \item \texttt{base learning}: A series of temporary $\theta$ parameters labelled as $\theta'$ are obtained via gradient descent as
  \begin{equation}\label{eq:GD1}
    \theta' \leftarrow \theta - \beta \nabla_\theta \mathcal{L}_\mathcal{T}(\Theta, \theta, S, \bar{S}),
  \end{equation}
  where $\beta$ is the learning rate in the base learning step and $\mathcal{L}_\mathcal{T}$ the loss function. The training is performed over only 3 epochs to prevent overfitting.
  \item \texttt{meta-learner update}: The $\theta$, scaling and shifting parameters are updated by one step of gradient descent as
  \begin{equation}\label{eq:GD2}
    \begin{aligned}
    \theta &=: \theta - \gamma \nabla_\theta \mathcal{L}_\mathcal{T}(\Theta, \theta', S, \bar{S}),\\
    S &=: S - \gamma \nabla_S \mathcal{L}_\mathcal{T}(\Theta, \theta', S, \bar{S}),\\
    \bar{S} &=: \bar{S} - \gamma \nabla_{\bar{S}} \mathcal{L}_\mathcal{T}(\Theta, \theta', S, \bar{S}),
    \end{aligned}
  \end{equation} 
  where $\gamma$ is the learning rate in the meta-learner updating step. After completing this step, the temporary parameters $\theta'$ will not be used anymore and can be discarded.
  \item \texttt{evaluation of $\mathcal{L}_{\mathcal{T}}$}: The loss function is evaluated using the updated parameters: $\mathcal{L}_\mathcal{T}(\Theta, \theta, S, \bar{S})$. This will be used to determine when to stop meta-training.
\end{itemize}
During the base learning and meta-learner update, the NN is trained to distinguish pure samples of 2.5k signal and 2.5k background from the SR region. A fifth of the sample is used for validation, which is also performed on pure samples. Different events are used for each of the three steps in the inner-loop of every episode. Once the inner-loop is complete, the $\mathcal{L}_{\mathcal{T}}$ are averaged and used to test for early stopping. Once the meta-training phase has been completed, the $\theta$ are reinitialized randomly. 

Third, fine-tuning is performed in an almost identical manner to Sec.~\ref{Sec:TransferLearning}. The only difference is now the presence of the scaling and shifting parameters that are learned during meta-training but kept fixed in this phase.

We mention that our method is simplified with respect to the original method of Ref.~\cite{sun2019meta}. The main difference was that we dropped the use of the hard tasks algorithm, as we considered this beyond the scope of a first study on the applicability of meta-learning to CWoLa. We also did not implement meta-batches, the meta-learning equivalent of a batch, as this was mostly irrelevant without the hard tasks algorithm.

\begin{figure}[t!]
    \centering
    \begin{subfigure}{0.49\textwidth}
        \centering
        \includegraphics[width=\textwidth]{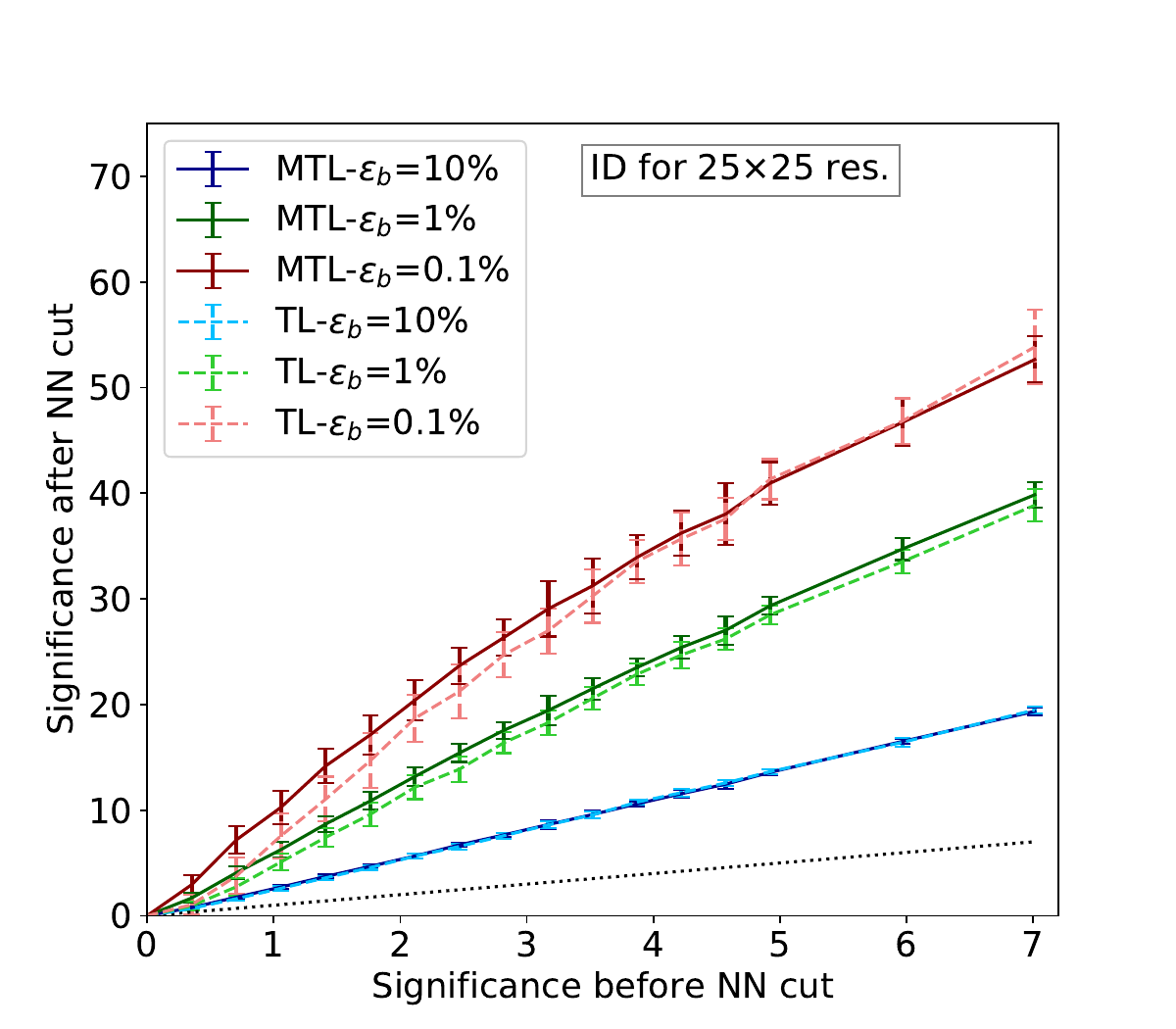}
        \caption{}
    \end{subfigure}    
    \begin{subfigure}{0.49\textwidth}
        \centering
        \includegraphics[width=\textwidth]{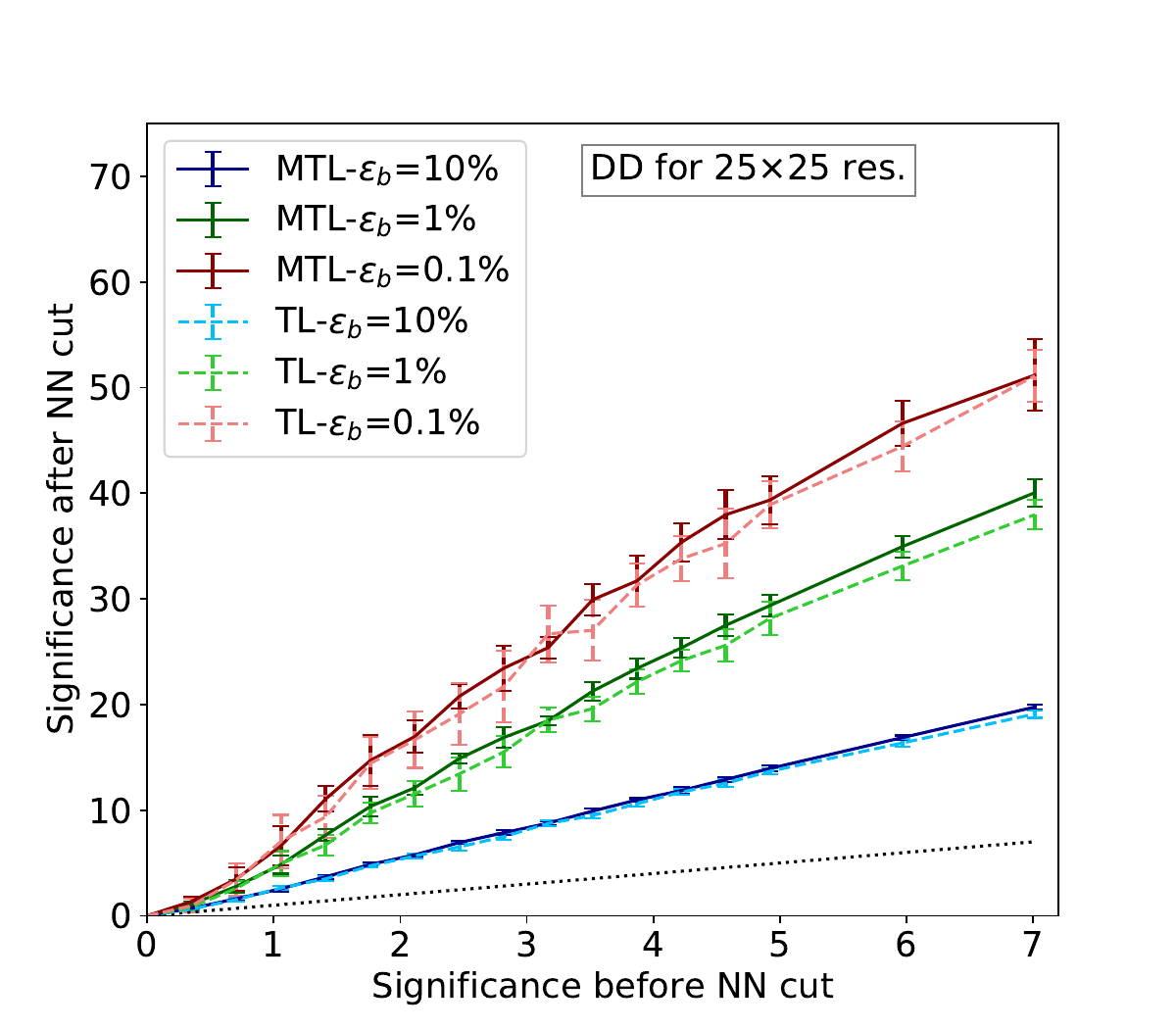}
        \caption{}
    \end{subfigure}    

    \begin{subfigure}{0.49\textwidth}
        \centering  \includegraphics[width=\textwidth]{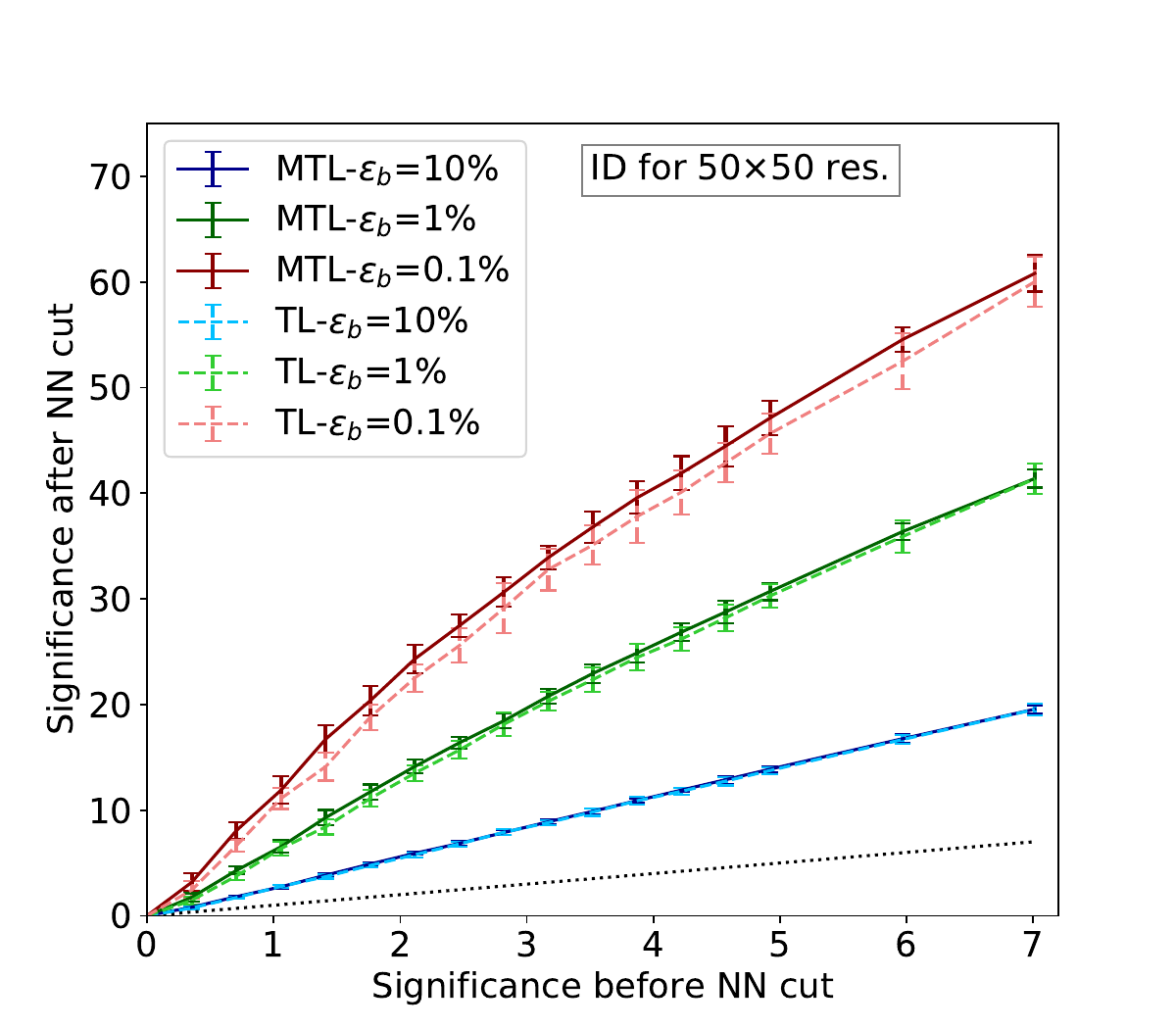}
        \caption{}
    \end{subfigure}    
    \begin{subfigure}{0.49\textwidth}
        \centering
        \includegraphics[width=\textwidth]{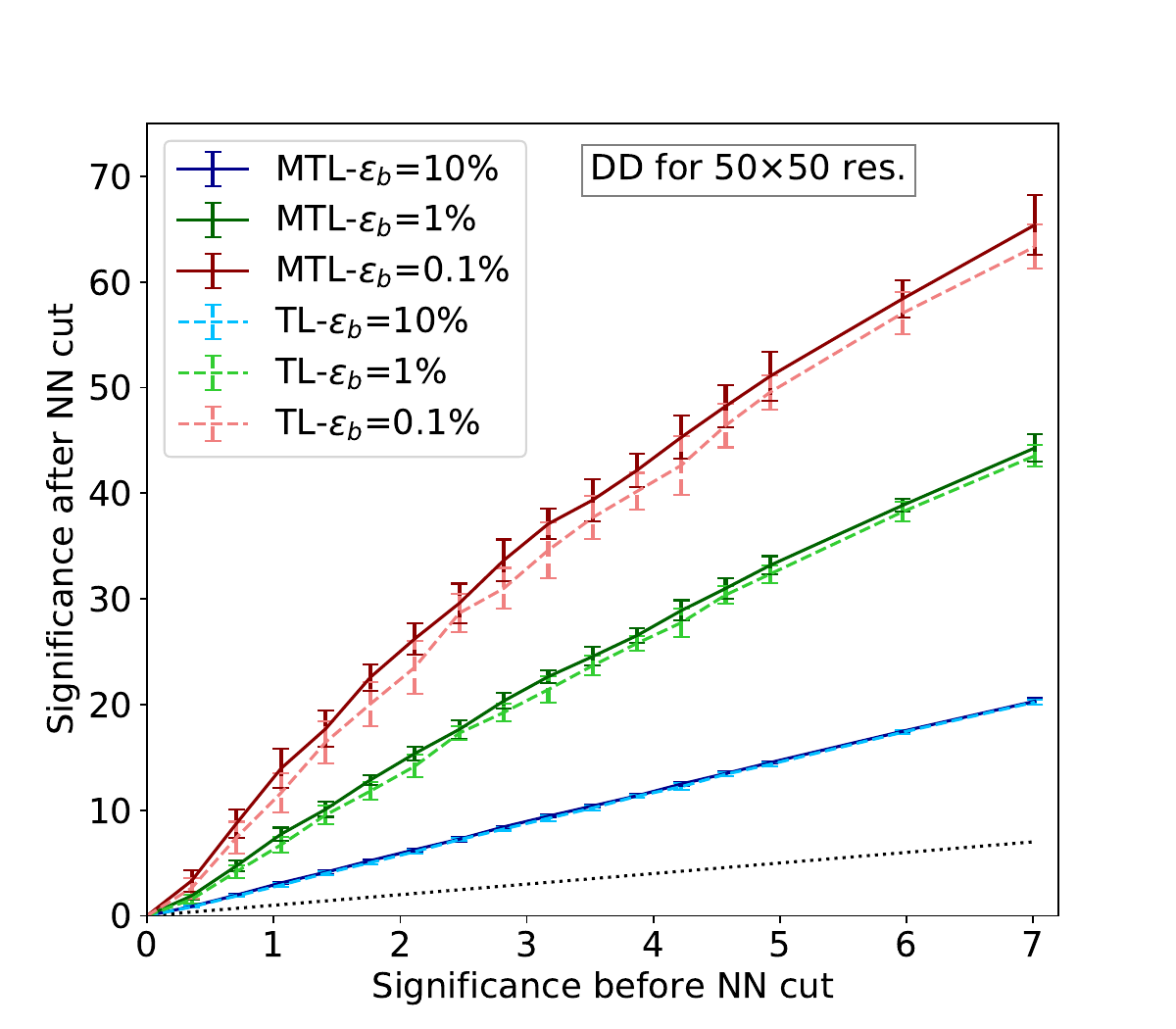}
        \caption{}
    \end{subfigure}  

    \caption{The results of meta-transfer learning (solid curves) and transfer learning (dashed curves, same as those in Fig.~\ref{fig:TL_CNN}) for the ID (left column) and DD (right column) scenarios with $\Lambda_D$= 10~GeV for $25\times25$ and $50\times50$ resolutions.  The dotted line in each plot has a slope of 1.}
    \label{fig:MTL_TL_CNN}
\end{figure}

\begin{figure}[t!]
    \centering
    \begin{subfigure}{0.49\textwidth}
        \centering
        \includegraphics[width=\textwidth]{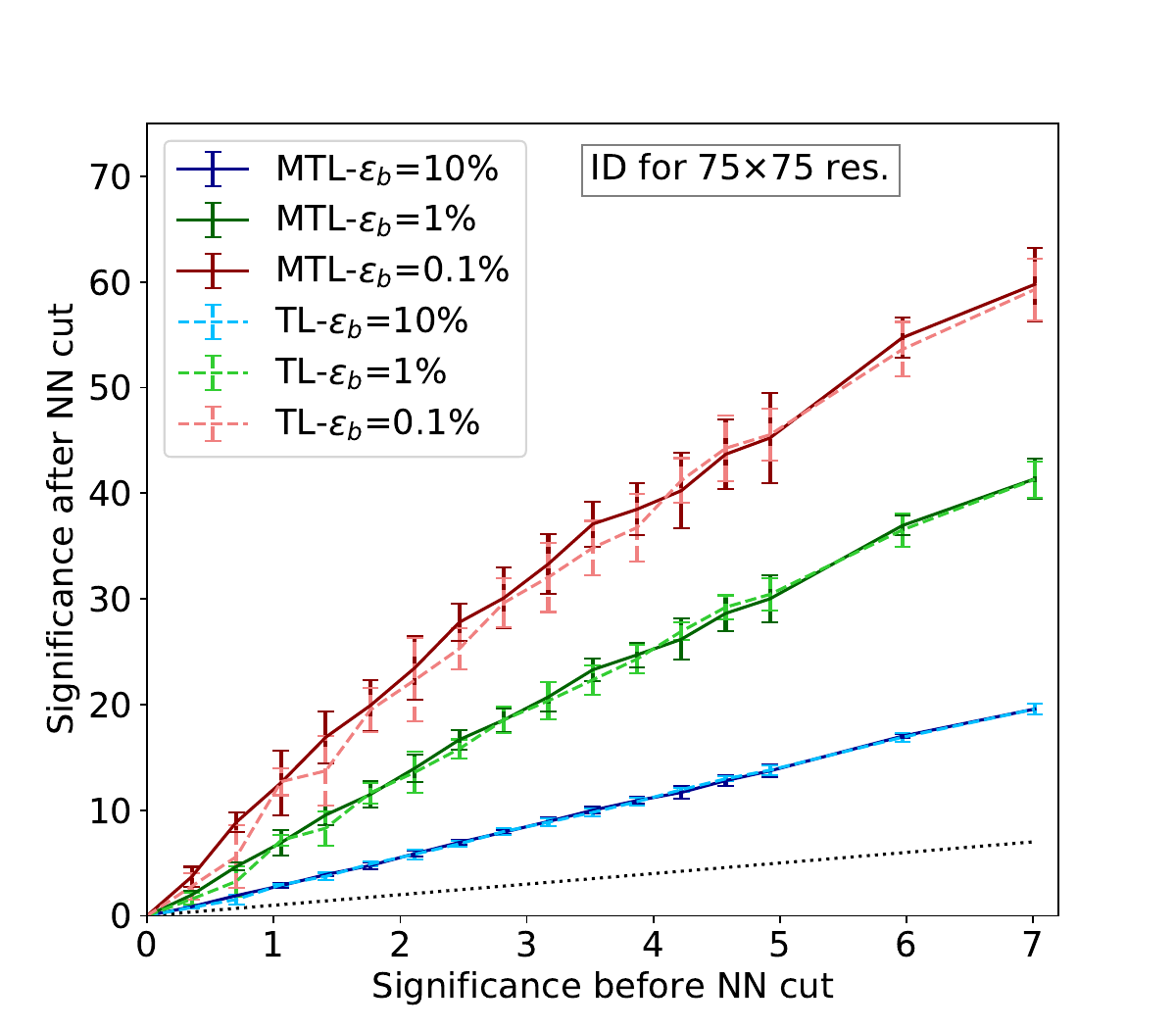}
        \caption{}
    \end{subfigure}    
    \begin{subfigure}{0.49\textwidth}
        \centering
        \includegraphics[width=\textwidth]{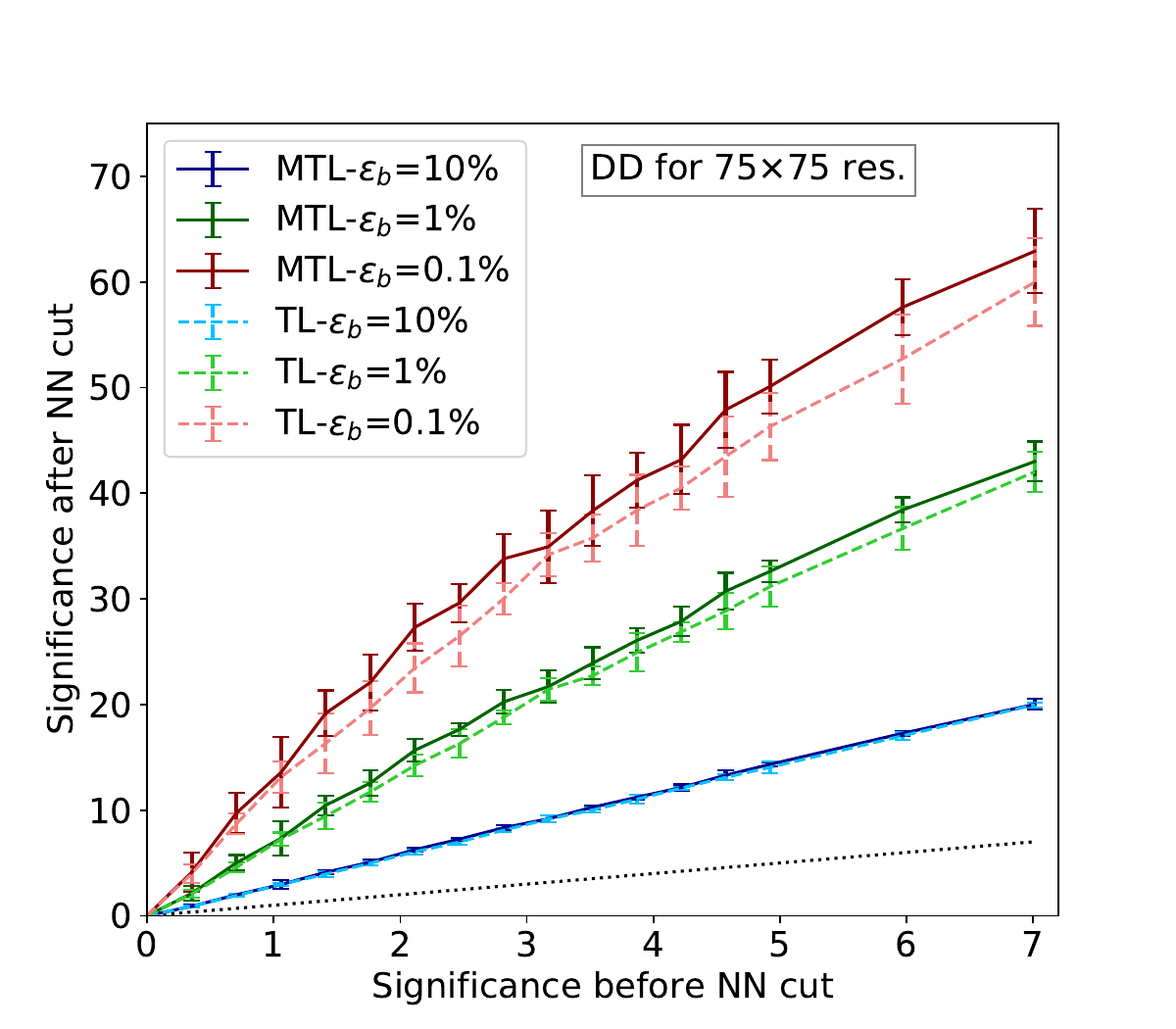}
        \caption{}
    \end{subfigure}    
    \caption{The results of meta-transfer learning (solid curves) and transfer learning (dashed curves) for the ID (left) and DD (right) scenarios with $\Lambda_D$= 10~GeV for $75\times75$ resolution with a larger size of kernels. The kernel sizes are $10\times10$ and $5\times5$ respectively instead of $5\times5$ and $3\times3$ mentioned in Table~\ref{tab:CNN_set}.  The dotted line in each plot has a slope of 1.}
    \label{fig:MTL_TL_CNN-large}
\end{figure}

Fig.~\ref{fig:MTL_TL_CNN} shows the comparison between transfer learning and meta-transfer learning. First, meta-transfer learning displays mostly a slight improvement in performance for the $25\times25$ and $50\times50$ resolutions compared with transfer learning due to the additional adjustment provided by the scaling and shifting. Note that the results from transfer learning are already good enough that mathematically there is not much room for improvement at large significance. The relative improvement at low significance can however be sizable. The difference between transfer and meta-transfer learning is negligible for the $75\times75$ resolution. However, we find that meta-transfer learning can display superior results to transfer learning even for the $75\times75$ resolution if a larger kernel size is used, and Fig.~\ref{fig:MTL_TL_CNN-large} shows the comparison. A full study of this is beyond the scope of this work though.

\section{Conclusion}\label{Sec:Conclusion}

Weak supervision searches have the advantages of both being able to train on data and being able to exploit distinctive signal properties. However, training a neural network via weak supervision can require a prohibitive amount of signal, often close to the point that the signal would already have been discovered without employing the neural network. In our work, we seek to address this problem by creating neural networks that can learn from less signal by using transfer and meta-learning. The general idea is to first train a neural network on simulations. During this stage, the neural network should learn relevant concepts or become a more efficient learner. The neural network is then trained on the experimental data and should require less signal because of its previous training. Our actual implementation of this procedure was via pretraining and meta-transfer learning.

We find that transfer learning can drastically improve the performance of CWoLa searches. The improvement is most important at low significance and the amount of signal necessary for discovery can be reduced by a factor of a few. Meta-transfer learning can further improve the performance of CWoLa searches, but not drastically so.

We mention that this work was intended more as a proof of principle and that there are still questions that are left unanswered. Namely, the choice of models on which to train could potentially have an effect on the ability to discover signals that differ considerably from them. The exact extent of this effect is left for future work. However, a small reduction to the scope of model sensitivity seems a fair prize to pay for the magnitude of our improvement over the regular CWoLa method.

Finally, we emphasize that transfer and meta-learning are vast and fast-evolving fields. Though we did demonstrate their potential, we only considered two specific techniques. It is plausible that more powerful techniques already exist or, even more likely, could be created in the future. Likewise, we have not sought to fully optimize our analysis and it is clear that some smaller details could be improved. As such and considering our very promising results, we believe that further studies of transfer and meta-learning for weak supervision are warranted.

\acknowledgments
We thank Chun-Hung Lin for collaboration in the early stage of this work. This work was supported by the National Science and Technology Council under Grant No. NSTC-111-2112-M-002-018-MY3, the Ministry of Education (Higher Education Sprout Project NTU-112L104022), and the National Center for Theoretical Sciences of Taiwan.

\appendix

\section{Impact of systematic uncertainties}\label{appendix:systematic-uncertainty}
A legitimate question is whether systematic uncertainties, which were not included up to now, can affect whether transfer and meta-learning provide an improvement over regular CWoLa. Although a full study is beyond the scope of this work, we try to answer this question in this appendix.

The only practical difference is that the significance should now be computed using the formula~\cite{ATLAS:2020yaz}
\begin{align}\label{eq:significance-uncertainty}
\sigma=\sqrt{2\left(N\log\left(\frac{N(N_b+\sigma^2)}{N_b^2+N\sigma^2}\right)-\frac{N_b^2}{\sigma^2}\log\left(1+\frac{\sigma^2 N_s}{N_b(N_b+\sigma^2)}  \right)\right)},
\end{align}
where $N_s$ and $N_b$ are respectively the numbers of signal and background, $N=N_s+N_b$ is the total number of events and $\sigma$ is the background systematic uncertainty. 

\begin{figure}[t!]
    \centering
      \includegraphics[width=0.5\textwidth]{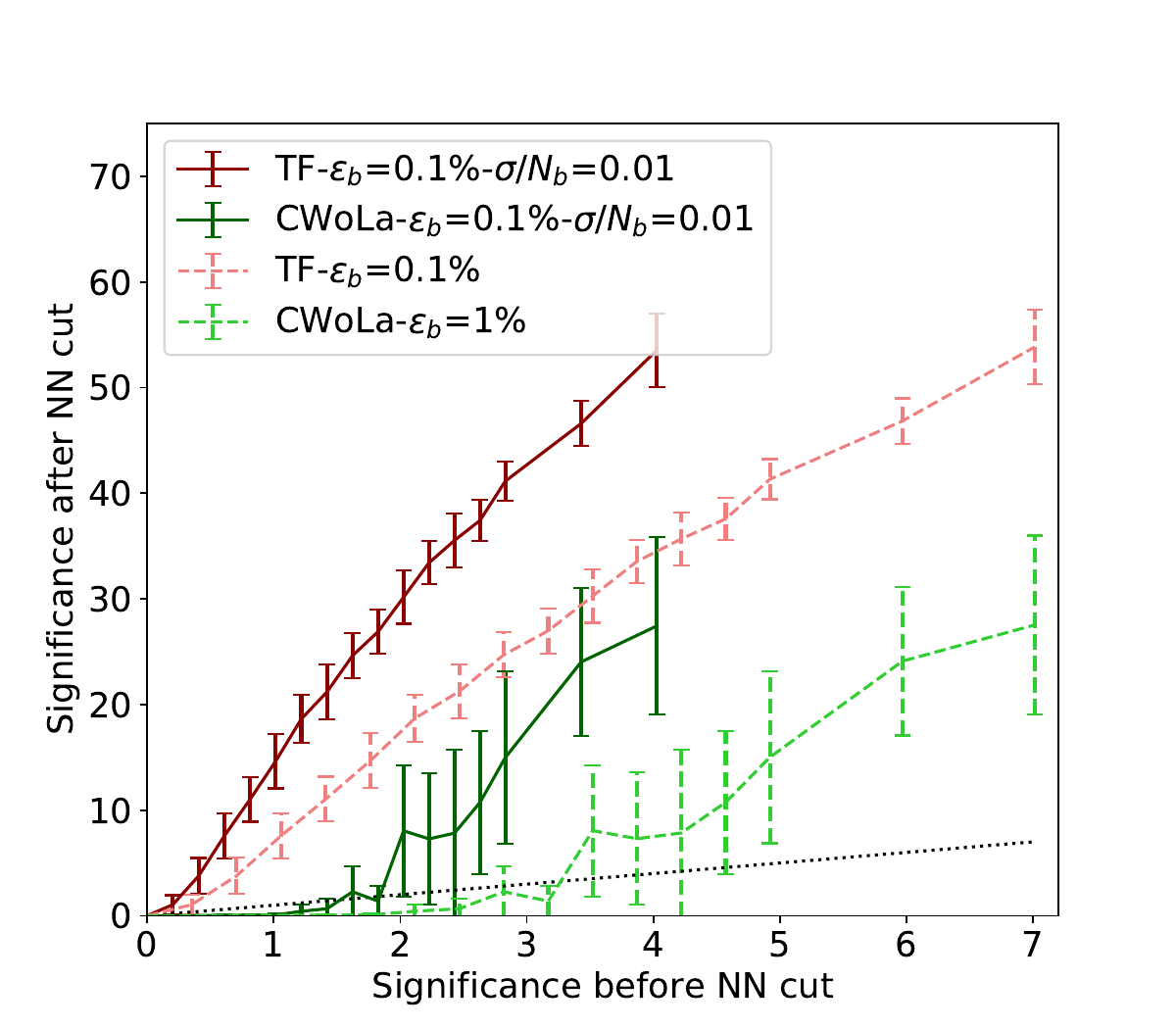}
    \caption{The results of transfer learning and pure CWoLa for the ID scenarios with $\Lambda_D$= 10~GeV for $25\times25$ resolutions with a systematic uncertainty of $1\%$. The solid curves are the results with systematic uncertainty and the dashed curves are the same as in Fig.~\ref{fig:TL_re25_A1}. The dotted line has a slope of 1.} 
    \label{fig:significance-uncertainty}
\end{figure}

Fig.~\ref{fig:significance-uncertainty} illustrates the impact of systematic uncertainties. Results for regular CWoLa and transfer learning are included. A relative systematic uncertainty of 1\% is assumed for the number of background events both before and after the NN cuts. This choice is made for illustration purposes, as it is somewhat smaller than the typical current precision of about 5\% for these masses, see e.g., Ref.~\cite{CMS:2019gwf}. Essentially, the inclusion of systematic uncertainties simply compresses the curves along the horizontal direction. This effect is accentuated as the systematic uncertainties increase. The reason is that systematic uncertainties reduce far more the significance before the NN cut than after. Unless the systematic uncertainties associated with transfer or meta-learning are far larger than those associated with CWoLa, transfer and meta-learning will therefore still outperform regular CWoLa.

\bibliography{biblio}
\bibliographystyle{utphys}

\end{document}